\documentclass[hyper]{JHEP3}

\newcommand{\Scal}[1]{\Bigl ({#1} \Bigr )}
\newcommand{\scal}[1]{\bigl ({#1} \bigr )}
\def\ie{{\it i.e.}\ }
\def\N{\mathcal{N}}
\DeclareMathAlphabet{\mathpzc}{OT1}{pzc}{m}{it}
\usepackage{dsfont}

\newcommand{\ord}[1]{{\scriptscriptstyle (#1)}}

\def\gl{\mathfrak{gl}}
\def\g{\mathfrak{g}}
\def\k{\mathfrak{k}}
\def\sl{\mathfrak{sl}}

\newcommand{\CR}{\nonumber \\*}
 \usepackage{amssymb}
 \usepackage{amsmath}

\renewcommand\Im{\hbox{{\rm Im}}\,}
\renewcommand\Re{\hbox{{\rm Re}}\,}
\def\cF{{\mathcal F}}
\def\cH{{\mathcal H}}
\def\cJ{{\mathcal J}}
\def\cR{{\mathcal R}}
\def\cV{{\mathcal V}}
\def\cK{{\mathcal{K}}}
\newcommand{\Iprod}[2]{\langle {#1}, {#2} \rangle}

\newcommand{\ft}[2]{{\textstyle\frac{#1}{#2}}}

\title{Duality covariant non-BPS first order systems}

\author{Guillaume Bossard  and Stefanos Katmadas
\\ {Centre de Physique Th\'eorique, \'Ecole Polytechnique, CNRS,\\ 91128
Palaiseau, France}
\\ \email{guillaume.bossard [at] cpht.polytechnique.fr},\\
   \email{stefanos.katmadas  [at] cpht.polytechnique.fr}}

\abstract{
  We study extremal black hole solutions to four dimensional $\N=2$  su\-per\-gra\-vi\-ty based on a cubic symmetric scalar manifold. Using the coset construction available for these models, we define the first order flow
 equations implied by the corresponding nilpotency conditions on the three-dimensional scalar momenta for the
 composite non-BPS class of multi-centre black holes.
 As an application, we directly solve these equations for the single-centre
 subclass, and write the general solution in a manifestly duality covariant form. This includes all single-centre
 under-rotating non-BPS solutions, as well
 as their non-interacting multi-centre generalisations.
}
\preprint{ 
           {CPHT-RR024.0512} }

\keywords{Black Holes in String Theory, Supergravity Models}

\begin{document}

\section{Introduction and Overview}
\label{sec:intro}

Black holes in the context of string theory have been a long lasting field of research, due to its deep connection to fundamental aspects of the theory. One important facet of the subject has been the construction of supergravity solutions describing the low energy strong coupling regime of these systems. For supersymmetric black holes, it has been possible to find these solutions explicitly, making use of the constraint imposed by the residual supercharges, even including higher derivative corrections, see {\it e.g.}~\cite{Ferrara:1995ih, Strominger:1996kf, Ferrara:1996dd, Behrndt:1997ny, Denef:2000nb, LopesCardoso:2000qm, Gauntlett:2002nw, Gauntlett:2004qy, Castro:2008ne}. Such a task however has proven more difficult for more general black holes, since one is forced to consider the full second order equations of motion rather than the first order BPS conditions. The simplest generalisation is the class of extremal black holes, which are still characterised by a vanishing Hawking temperature, but do not 
preserve any supersymmetry. The corresponding static solutions are known to be described by first order equations as well \cite{Ceresole:2007wx,LopesCardoso:2007ky,Ceresole:2009iy,Gimon:2007mh,Gaiotto:2007ag,Andrianopoli:2007gt,Andrianopoli:2009je,Bossard:2009we,Ceresole:2009vp,Perz:2008kh,Kim:2010bf,Galli:2010mg}, although the latter are then not a direct consequence of supersymmetry.

Extremal black holes in supergravity theories fall in two distinct categories, namely the under-rotating and the over-rotating branches, where the former contains several subclasses. The over-rotating (or ergo) branch is characterised by the presence of an ergo-region and includes the extremal Kerr solution. In contrast, we will focus on the under-rotating (or ergo-free) black holes, which then admit a flat three-dimensional base, and include the static extremal black holes \cite{Rasheed:1995zv, Matos:1996km, Larsen:1999pp}. Single centre under-rotating non-BPS black holes have been studied throughout the last decade or so, from various aspects and using various techniques, see for example  \cite{Ortin:1996bz, Khuri:1995xq, Tripathy:2005qp, Astefanesei:2006dd, Ceresole:2007wx, Andrianopoli:2007gt, Perz:2008kh, Ceresole:2009iy, Bossard:2009we, Ceresole:2009vp} and references therein for some developments. Using the seed solution of \cite{LopesCardoso:2007ky, Gimon:2007mh, Bena:2009ev} combined with a general 
duality transformation as explained in \cite{Dall'Agata:2010dy}, one can construct any desired solution, but a manifestly duality covariant formulation was lacking. 

For theories coupled to a symmetric scalar manifold, as the ones we will deal with, there are three classes of solutions describing interacting ergo-free extremal black holes, distinguished by the algebraic properties of the first order systems that describe them. These are the standard BPS solutions \cite{Behrndt:1997ny, Denef:2000nb}, the so-called almost BPS solutions \cite{Goldstein:2008fq}, and the composite non-BPS solutions \cite{Bossard:2011kz}. After the BPS class, the almost-BPS class is perhaps the best studied, especially in the five dimensional uplift, where it was originally discovered and extended, see {\it e.g.}~\cite{Bena:2009ev, Bena:2009en, Bena:2009qv, Bena:2009fi, Bobev:2009kn}. On the other hand, the composite non-BPS class appears to be simpler in the four dimensional context and is the class we discuss in the following.

In this paper we will consider the equations of motion for stationary solutions as described by the dimensional reduction of the theory along time, \ie as a non-linear sigma model over a pseudo-Riemannian symmetric space coupled to three-dimensional Euclidean gravity \cite{Breitenlohner:1987dg}. Spherically symmetric black hole solutions correspond to geodesics on this symmetric space and, in particular, extremal static black holes correspond to the subclass of null geodesics. It was shown in \cite{Gunaydin:2005mx, Gaiotto:2007ag, Bergshoeff:2008be, Bossard:2009at, Bossard:2009my} that null geodesics associated to extremal black holes are classified by the  nilpotent orbits in which the corresponding Noether charges lie in, thus recasting the equations of motion to an eigenvalue equation for the momentum of the coset scalars. The classification of extremal solutions in terms of nilpotent orbits has been studied in details in \cite{Bossard:2009we,Kim:2010bf,Fre:2011uy,Chemissany:2012nb}. 

Considering a general stationary Ansatz with a flat three dimensional base, one finds that the only regular solutions are described by scalar fields taking values in a nilpotent subgroup of the three-dimensional duality group. Using the nilpotent orbits classification, one can determine pertinent nilpotent subalgebras, which lead to the description of stationary solutions describing interacting black holes. It was shown that all the three systems, BPS, almost BPS and composite non-BPS can be described in terms of associated nilpotent subalgebras \cite{Bossard:2011kz,Bossard:2012ge}. In these papers, explicit examples of these solutions were constructed in terms of specific nilpotent subalgebras.

However, these seed solutions are not duality covariant by construction, whereas one should be able to construct them without referring to a specific representative. Although the general solutions can be obtained from these seed solutions by duality transformations, the resulting form of the solutions is not easy to parametrize in terms of physically relevant quantities. A duality covariant formulation of these solvable systems would permit to obtain these solutions in a form exhibiting their physical properties. This is of particular importance when addressing issues such as existence and stability of composite bound states. For the BPS solutions (as well as the non-BPS $Z_*=0$), the first order system associated to the corresponding nilpotent orbits was written in a manifestly duality invariant form \cite{Bossard:2010mv}, leading to a generalisation of the BPS solutions \cite{Behrndt:1997ny, Denef:2000nb} to $\N=8$ supergravity, and non-BPS $Z_*=0$ solutions. 

In what follows, we explain how the composite non-BPS solutions can also be described in a manifestly duality invariant form. As it turns out, the system is characterised by a constant very small charge vector (\ie such that its quartic invariant satisfies $I_4 = \partial I_4 = \partial^2 I_4|_{\rm \scriptscriptstyle ad}= 0$), which has only one charge component in an appropriate duality frame. This represents an auxiliary variable that restricts the types of charges allowed in the various centres, or conversely is fixed by the physical charges for a regular solution.

As a first application, we will solve explicitly the system in the restricted case describing single centre (or non-interacting multi-centre) black holes in a manifestly duality covariant way. The reduction to this subclass is effected by a suitable reality constraint on the scalar momenta, which we use systematically to simplify the problem. The result is a stabilisation equation for the scalars throughout the black hole background that parallels the one derived in \cite{Behrndt:1997ny, Denef:2000nb} for supersymmetric solutions. However, we find new terms, proportional to the very small vector driving the flow and its magnetic dual, with coefficients depending on the harmonic functions carrying the electromagnetic charges and angular momentum. These represent a duality covariant realisation of the simple change of sign for a particular charge, which has been observed to relate some simple non-BPS solutions to supersymmetric ones \cite{Khuri:1995xq,Tripathy:2005qp,Kallosh:2006ib}, and the non-harmonic term 
obtained in \cite{Galli:2010mg} by considering the seed solution of \cite{Bena:2009ev}, respectively.

Understanding how to solve the first order system when restricted to single centre solutions is a necessary step towards the resolution of the composite non-BPS system. In order to do so, one must rewrite the non-linear first order system we describe in this paper, in a second order linear system of differential equations. The explicit solution of the composite non-BPS system will be the purpose of forthcoming research. 

We start by setting up notation and giving some background on both four dimensional $\N=2$ supergravity and the corresponding non-linear sigma model obtained by time-like reduction in section \ref{sec:coset}. In section \ref{sec:composite-syst} we proceed to the discussion of the algebraic structure of the composite non-BPS system and discuss the reality constraint that reduces it to describe single centre solutions. We then go on to solve explicitly the flow equations for the single centre class and discuss some aspects of the solutions obtained in section \ref{sec:single-class}. We conclude in section \ref{sec:concl} with some general remarks and future plans of extending to the multi centre classes, whereas the two appendices are devoted to the illustrating the general solution with the seed solution  of \cite{Bena:2009ev} and to the derivation of the duality invariant constraints on the charges.

\section{Non-linear sigma model formulation of stationary solutions}
\label{sec:coset}
In this section, we collect various formulae and conventions which will be
essential for the connection of the objects appearing in the three-dimensional non-linear sigma model formulation of  $\N=2$ supergravity describing stationary solutions to standard four-dimensional supergravity variables.

\subsection{$\N=2$ supergravity and symmetric special K\"{a}hler geometry}\label{sec:sugra}
The bosonic Lagrangian of $\N=2$ supergravity coupled to $n_v$ vector multiplets
reads \cite{deWit:1984pk, deWit:1984px}
\begin{eqnarray}\label{eq:Poincare-4d}
8\pi\,e^{-1}\, {\cal L} &=&
  - \ft12  R -i \, \Iprod{D^{\mu} \bar \cV}{D_{\mu} \cV}
-\ft{\mathrm{1}}4\, F^I_{\mu\nu}\, G_I^{\mu\nu} \,,
\end{eqnarray}
where the $F^{I}_{\mu\nu} = \partial_\mu A^I_\nu - \partial_\nu A_\mu^I$ for $I=0,\dots n_v$ encompass the graviphoton and the
gauge fields of the vector multiplets and $G_I^{\mu\nu}$ are the dual field strengths, defined in terms of the $F^I_{\mu\nu}$ though the scalar dependent
couplings,
whose explicit form will not be relevant in what follows. The gauge field
equations of motion and Bianchi identities can then be cast as a Bianchi
identity on the symplectic vector
\begin{equation}\label{eq:dual-gauge}
 \cF_{\mu\nu}=\begin{pmatrix} F_{\mu\nu}^I\\ G_{I\, \mu\nu}\end{pmatrix}\,,
\end{equation}
whose integral over any two-cycle defines the associated electromagnetic charges
through
\begin{equation}
\Gamma=\begin{pmatrix} p^I\\ q_{I}\end{pmatrix}
 =\frac{1}{2\pi} \,\int_{S^2} \cF\,.
\end{equation}

The physical scalar fields $t^i$, which parametrize a special K\"ahler space $\mathcal{M}_4$ of complex dimension $n_v$, only appear in \eqref{eq:Poincare-4d} through the section, $\cV$, of a holomorphic $U(1) \times Sp(2n_v+2,\mathbb{R})$ bundle over $\mathcal{M}_4$.
Choosing a basis, this section can be written in components in terms of scalars
$X^I$ as
\begin{equation}\label{eq:sym-sec}
\cV=\begin{pmatrix} X^I\\ F_I\end{pmatrix}\,, \qquad
F_I= \frac{\partial F}{\partial X^I}\,,
\end{equation}
where $F$ is a holomorphic function of degree two, called the prepotential,
which we will always consider to be cubic
\begin{equation}\label{prep-def}
F=-\frac{1}{6}c_{ijk}\frac{X^i X^j X^k}{X^0} \equiv  -\frac{\N[X]}{X^0} \,,
\end{equation}
for completely symmetric $c_{ijk}$, $i=1,\dots n_v$, and we introduced the cubic
norm $\N[X]$. The section $\mathcal{V}$ is subject to the constraint
\begin{equation}
  \label{eq:D-gauge}
  \Iprod{\bar{\mathcal{V}}}{\mathcal{V}} =  i  \,,
\end{equation}
and is uniquely determined by the physical scalar fields $t^i$ up to a local $U(1)$ transformation. The $U(1)$ gauge invariance of \eqref{eq:Poincare-4d} is ensured by the appearance of the K\"{a}hler connection $Q_\mu$ in the covariant derivative. The K\"{a}hler potential on $\mathcal{M}_4$ is defined up to an arbitrary holomorphic function $f(t)$ as
\begin{equation} \cK = - \mbox{ln}\scal{   i \,  \N[t-\bar t]}  + f(t) + f( \bar t)  \end{equation}
and we fixed the $U(1)$ gauge invariance in terms of K\"ahler transformations by requiring that the K\"{a}hler connection is determined by the K\"{a}hler potential as
\begin{equation} Q = \Im[ \partial_ i \cK dt^i] \label{Kah-conn}\ ,\end{equation}
such that
\begin{equation}
 \label{Kah-metr-der}
g_{i\bar \jmath}= \partial_i\partial_{\bar \jmath} \cK\,, \qquad
D_\mu\cV=(\partial_\mu +i\,Q_\mu)\cV = D_i \cV \, \partial_\mu t^i
  = (\partial_i \cV + \tfrac{1}{2}\partial_i \cK\,\cV) \, \partial_\mu t^i\,,
\end{equation}
where $D_i\cV$ is the corresponding K\"ahler covariant derivative on the components of the
section. With the prepotential \eqref{prep-def}, the special geometry identities \cite{Ceresole:1995ca} reduce to 
\begin{equation}
 \bar D_{\bar \jmath} D_i \cV=\,g_{i\bar \jmath} \cV\,, \qquad
  D_i D_{j} \cV= i  e^{\cK} \,c_{i j k} g^{k\bar k} \bar D_{\bar k} \bar \cV\,, \label{SpecialGeoId} 
\end{equation}
which will be used extensively in what follows.

We introduce the following notation for any symplectic vector $\mathcal{J}$
\begin{align}
Z(\mathcal{J}) = \Iprod{\mathcal{J}}{\cV} \,,\\
Z_i(\mathcal{J}) = \Iprod{\mathcal{J}}{D_i \cV} \,,
\end{align}
with the understanding that when the argument is form valued, the operation is
applied component wise. For instance, the central charge of the gauge field is\begin{equation} Z(\cF)
 = e^{\frac{\cK}{2}} \Scal{ G_0 + t^i G_i +
\frac{1}{2} c_{ijk} t^i t^j F^k - \N[t] F^0 }\,, \end{equation}
for the prepotential \eqref{prep-def}. With these definitions it is possible to introduce a scalar dependent complex
basis for symplectic vectors, given by $(\cV,\, D_i\cV)$, so that any vector
$\cJ$ can be expanded as
\begin{equation}\label{Z-expand}
\cJ = 2 \Im[- \bar{Z}(\cJ)\,\cV + g^{\bar \imath j} \bar{D}_{\bar \imath}  \bar{Z}(\cJ)\, D_j \cV]\,,
\end{equation}
whereas the symplectic inner product can be expressed as
\begin{equation}\label{inter-prod-Z}
\Iprod{\cJ_1}{\cJ_2} = 2 \Im[- Z(\cJ_1)\,\bar{Z}(\cJ_2)
   + Z_a(\cJ_1) \, \bar{Z}^a (\cJ_2)]\,.
\end{equation}
Finally, we introduce the notion of complex selfduality of the gauge fields
\eqref{eq:dual-gauge}, which satisfy the identity
\begin{equation}
 \mathrm{J}\,\cF=-*\cF\,,\label{cmplx-sdual}
\end{equation}
where $\mathrm{J}$ is a scalar dependent complex structure defined as
\begin{equation}\label{CY-hodge}
\mathrm{J}\cV=-i \cV\,,\quad
\mathrm{J} D_i\cV=i  D_i\cV\,.
\end{equation}

 In this paper we will consider that $\mathcal{M}_4$ is moreover a symmetric space,
such that the coefficients $c_{ijk}$ are left invariant by the action of a group $G_5$.  In this case one can define the vielbeins on $\mathcal{M}_4$ such that
\begin{equation}  g^{i\bar \jmath}   = e_a^i e^{a\, \bar \jmath} \label{eq:kah-viel}, \end{equation}
and
\begin{equation} c_{abc} =  i  \,  e^{\cK} e_a^i e_b^j e_c^k\,  c_{ijk}, \label{Ctangent} \end{equation}
where $c_{abc}$ is a constant symmetric tensor left invariant by the action of $K_4$, which is the compact real form of $G_5$.  Then, the contravariant symmetric tensor $c^{abc}$ in the conjugate representation satisfies the Jordan identity \cite{Gunaydin:1983bi}
\begin{equation} c_{f(ab} c_{cd)g} c^{efg}  = \frac{4}{3} \delta^e_{(a} c_{bcd)}\,.
 \label{symmetric}\end{equation}
The symmetric space $\mathcal{M}_4$ is defined as the coset space of the four-dimensional duality group $G_4$ by its holonomy subgroup
\begin{equation} \mathcal{M}_4\cong \scal{ U(1)\times K_4} \big \backslash G_4 \, .\label{4d-coset} \end{equation}
The scalar fields can then equivalently be described by a coset representative $\upsilon$ in $G_4$, and the associated Maurer--Cartan form in the coset component $\mathds{C}^{n_v} \cong \mathfrak{g}_4 \ominus ( \mathfrak{u}(1) \oplus \k_4 )$ is defined as
\begin{equation} d \upsilon \upsilon^{-1} + \upsilon^{\dagger -1} d \upsilon^\dagger =  e^a_i d t^i \,   {\bf Y}_a  + e_{a \bar \imath} d \bar t^{\, \bar \imath} \, {\bf Y}^a \end{equation}
where ${\bf Y}_a$ (and Hermitian conjugate ${\bf Y}^a$) define a basis in  $\mathfrak{g}_4 \ominus ( \mathfrak{u}(1) \oplus \k_4 )$. Combined with \eqref{Kah-metr-der}, this equation permits to relate the expression of the Maurer--Cartan form to the derivative of the section $\cV$ in the tangent frame. Similarly, we also define
\begin{equation} Z_a(\cF) = e_a^i D_i Z(\cF) = \Iprod{\cF}{e_a^i D_i\cV}\, , \end{equation}
which transforms in the same $\mathds{C}^{n_v}$ representation of $K_4$. 

\subsection{Time-like reduction and para-quaternionic geometry}
In order to describe stationary asymptotically flat extremal black holes, we
introduce the standard Ansatz for the metric
\begin{equation}
 \label{metricMltc}
 ds^2=- e^{2U}( d t+\omega )^2 + e^{-2U}   d {\bf x} \cdot  d{\bf x} \,,
\end{equation}
in terms of a scale function $U(x)$ and the Kaluza--Klein one-form
$\omega(x)$ (with spatial components only), which are both required to asymptote to zero at
spatial infinity. Here and henceforth, all quantities are independent of
time, so that all scalars and forms are defined on the flat three-dimensional
base. The gauge fields are decomposed in a similar fashion as
\begin{equation} 2  \mathcal{A} = \zeta ( d t+\omega) + w \end{equation}
and accordingly for the field strengths
\begin{equation}
 \label{gauge-decop}
2  \cF= d \zeta \, ( d t+\omega) + F   \,, \qquad
   F= \zeta\, d  \omega + d w \,,
\end{equation}
where we defined the gauge field scalars $\zeta$, arising as the time component
of the gauge fields, and the one-forms $ w$ describing the charges. Here, $F$ is defined as the spatial component of the field strength, and is not closed but satisfies
\begin{equation} d F = e^{2U} \star d \omega \wedge J F \; , \label{F-closed}  \end{equation}
according to \eqref{cmplx-sdual}, which reads
\begin{equation}\label{cmplx-self}
 d \zeta = e^{2U}  \mathrm{J}\star F \,.
\end{equation}
Note that this first order equation determines the $\zeta$ in terms of the vector fields $w$ and the
scalars.

The scalar field $\sigma$ dual to the Kaluza--Klein vector $\omega$
\begin{equation} e^{4U} \star d \omega  = d \sigma + \langle \zeta , d \zeta \rangle \,,  \end{equation}
defines the coordinate of an $S^1$ fiber over the symplectic torus $\mathbb{T}$ parametrized by the $\zeta$'s. Altogether with the scaling factor $U$ and the moduli $t^i$, these fields parametrize the para-quaternionic symmetric  space \footnote{Here para-quaternionic refers to the property that the holonomy group of $\mathcal{M}_3$ $SL(2) \times G_4 \subset SL(2) \times Sp(2n_v,\mathds{R})$.}
\begin{equation} \label{M3}
\mathcal{M}_3 \cong G_3 / \scal{ SL(2) \times G_4 }  \ .
\end{equation}
This defines the so-called $c^*$-map, which can be related to the standard $c$-map \cite{Ferrara:1989ik} by analytic continuation. 
The three-dimensional symmetry group Lie algebra $\g_3$ decomposes as
\begin{equation} \label{3d-alg-decomp}
\mathfrak{g}_3 \cong \mathbf{1}^\ord{-2}\oplus \mathfrak{l}_4^\ord{-1}
  \oplus (\gl_1 \oplus \mathfrak{g}_4)^\ord{0}
  \oplus \mathfrak{l}_4^\ord{1} \oplus \mathbf{1}^\ord{2}\,,
\end{equation}
where the weights refer to the
eigenvalues under the adjoint action of the $\gl_1$ generator. The grade one generators in $\mathfrak{l}_4^\ord{1}$ are associated to the gauge invariance with respect to a constant shift of the scalars $\zeta$, and accordingly the grade two generator correspond to the invariance under shift of the scalar $\sigma$.

At this stage it is important to introduce some properties of the $\mathfrak{g}_3$ algebra. The components of an element of the Lie algebra $\g_3$ in the coset component $\g_3 \ominus ( \sl_2 \oplus \g_4) $, can be decomposed in terms of $U(1) \times K_4 $ irreducible representations as two  complex parameters ${\rm w}$ and $Z$ and two complex vectors $\bar Z^a , \Sigma^a$ which transform in the $\mathds{C}^{n_v}$ representation of $K_4$ (the same as the scalar fields momenta $e^a_i d t^i$ in four dimensions) such that $\N[\bar Z] \equiv \frac{1}{6} c_{abc} \bar Z^a \bar Z^b \bar Z^c$ is $K_4$ invariant. The quadratic trace invariant defines the $SL(2) \times G_4$ invariant norm
\begin{equation} | {\rm w} |^2 - |Z|^2  - Z_a \bar Z^a + \Sigma^a \bar \Sigma_a \label{Tracesquare}\ , \end{equation}
and the $\sl_2$ algebra is realised on these components as
\begin{equation} \delta  {\rm w} =i \rho  {\rm w}  + \bar \lambda Z\,, \quad 
\delta Z =-  i \rho Z  +  \lambda {\rm w}\,,  \quad 
\delta \bar Z^a = i \rho \bar Z^a + \bar \lambda \Sigma^a\,, \quad 
\delta \Sigma^a = -i \rho \Sigma^a + \lambda \bar Z^a\,,
\label{sl2-alg}\end{equation}
for a complex $\lambda$. Closure of this algebra can conveniently be checked in the Cartan complex, considering
anticommuting parameters $\lambda, \rho$ and the differential $\delta$
\begin{equation} \delta \rho = i \lambda \bar \lambda\,, \qquad 
\delta \lambda = - 2 i \rho \lambda\, , \end{equation}
whose nilpotency is equivalent to the Jacobi identity, similar to the BRST formalism in gauge theories. In the same way, the $\g_4$ algebra is realised in terms of
the elements of $\k_4$, denoted by $G^a{}_b$, a real $\gamma$ and a complex vector
$\Lambda_a$, associated to the decomposition
\begin{equation} \g_4 \cong \mathfrak{u}(1) \oplus \k_4
\oplus \mathds{C}^{n_v}\,, \end{equation}
which defines the symmetric space $\mathcal{M}_4$. The action of $\g_4$ on the coset component can be written as
\begin{equation}\begin{split} \delta {\rm w} &= \Lambda_a \bar Z^a + 3 i \gamma {\rm w}\\
\delta Z &= \Lambda_a \Sigma^a + 3 i \gamma Z  \end{split}\qquad
\begin{split} \delta \Sigma^a &= \bar \Lambda^a Z + c^{abc} \Lambda_b Z_c +
G^a{}_b \Sigma^b + i \gamma \Sigma^a \ , \\
\delta \bar Z^a &=  \bar \Lambda^a {\rm w}  + c^{abc} \Lambda_b \bar \Sigma_c +
G^a{}_b\bar Z^b + i \gamma \bar Z^a \ .  \end{split}\end{equation}
The corresponding algebra is realised in terms of anticommuting parameters with the nilpotent differential
\begin{align}
\delta \Lambda_ a = - G^b{}_a \Lambda_b + 2 i \gamma \Lambda_a
\qquad & \delta \gamma = \frac{i}{3} \bar \Lambda^a \Lambda_a \ , \CR
\delta G^a{}_b = G^a{}_c
G^c{}_b + c^{ace} c_{bde} \Lambda_c \bar \Lambda^d +& \bar \Lambda^a \Lambda_b +
\frac{1}{3} \bar \Lambda^c \Lambda_c \delta^a_b \ . 
\end{align}
Note that the variation of $G^a{}_b$ indeed leaves invariant the cubic norm $\N[Z]$ for an anticommuting $\Lambda_a$.

We now consider the equations of motion for the scalar fields parametrizing the symmetric space $\mathcal{M}_3$. These are expressed in terms of the corresponding Maurer--Cartan form
\begin{equation} v^{-1} d v = P + B\,, \end{equation}
which decomposes accordingly into the coset component $P$ defining the scalar momenta, and the $\sl_2 \oplus \g_4$ component $B$ defining the pulled back spin connection. In components, the scalar momenta are defined as
\begin{equation} {\rm w} \equiv - d U - \frac{i}{2} e^{2U} \star d \omega \,, \quad \Sigma^a =- e^a_i d t^i\,, \quad 
Z \equiv e^U Z(\star F)\,, \quad Z_a \equiv e^U Z_a(\star F)\ , \end{equation}
where we introduce some shorthand notations that will be used for the remainder of the section.
Analogously, we give the components of $B$ along $\sl_2$
\begin{equation} \rho(B) = -\tfrac{1}{4} \mathrm{e}^{2U} \star d \omega - \tfrac{1}{2} Q \,, \qquad
\lambda(B) =  e^U Z(\star F)  \,, \label{Bsl2} \end{equation}
its components along $\g_4$
\begin{equation}
\gamma(B) = -\tfrac{1}{4} \mathrm{e}^{2U} \star d \omega + \tfrac{1}{6} Q \,, \qquad
\Lambda_a(B) = e^U Z_a(\star F) \,,  \label{Bg4} \end{equation}
and finally $G^a{}_b(B)$ defines the $\k_4$ valued traceless component of the pulled back spin connection on $\mathcal{M}_4$:\begin{equation} G^a{}_b(B) = e^a_i \partial_{\bar \jmath} e^i_b \, d \bar t^{\, \bar \jmath} - e_{\bar \jmath b} \partial_i e^{\bar\jmath a} d t^i - \frac{2i}{3} \delta^a_b Q \ , \label{Bk4} \end{equation}
where $Q$ is the pulled back K\"{a}hler connection \eqref{Kah-conn}.\footnote{To prove that $G^a{}_b(B)$ is indeed traceless, one can use \eqref{symmetric} to show that $c^{acd} c_{bcd} = \frac{n_v+3}{3} \delta^a_b$ and substitute \eqref{Ctangent} in $c^{abc} \bar \partial_{\bar \imath} c_{abc} = 0 $.}

These formulae given, one can straightforwardly compute the equations of motions of the scalar and vector fields respectively as coming from the equation of motion and Bianchi identity on $P$, as follows
\begin{equation} d_B \star P = 0\,,\qquad d_B P = 0\,, \label{eom-bian-coset}\end{equation}
where $d_B$ stands for the covariant derivative on the coset. For instance, the components of the equation of motion for $P$ are
\begin{eqnarray} d_B \star  {\rm w}  &=& - d\star d U - \frac{1}{2} e^{4 U } d \omega \star d \omega  + \ e^{2U} \mbox{Re}\bigl[ Z(F) \bar Z(\star F) + Z_a(F) \bar Z^a(\star F) \bigr] = 0 \ ,  \CR
d_B \star \Sigma^a &=& - \nabla \star e^a_i d t^i + e^{2U} \scal{ 2 Z(F) \bar Z^a(\star F) + c^{abc} Z_b(F) Z_c(\star F) } = 0  \ ,
\end{eqnarray}
and
\begin{eqnarray}  d_B \star Z &=& e^U \Scal{ D Z(F) - e_i^a dt^i \wedge Z_a(F) - i e^{2U} \star d \omega \wedge  Z(F)  } = 0 \; , \\
d_B \star Z_a &=& e^U \Scal{ D Z_a(F) - e_{a \bar \imath} d\bar t^{\, \bar \imath} \wedge  Z(F) - c_{abc} e^b_i dt^i \wedge \bar Z^c(F) + i e^{2U} \star d\omega \wedge Z_a(F) } = 0 \; , \nonumber
\end{eqnarray}
where \eqref{F-closed} and the standard special geometry identities \eqref{SpecialGeoId} were used. In the following we will not analyse these equations directly, but will rather employ arguments based on the nilpotency of $P$ for extremal solutions to obtain equivalent first order equations that can be solved directly.

\section{Nilpotent orbits and first order systems}
\label{sec:composite-syst}

In this section we generalise the formalism developed in \cite{Bossard:2010mv} to arbitrary nilpotent orbits of $G_3$, with a specific emphasis to the ones describing non-BPS black holes. The basic observation is that the only regular stationary solutions of $\N=2$ supergravity with a flat three-dimensional base metric are such that the momentum $P$ is nilpotent. This implies in particular that $P$ can be written in a basis of generators ${\bf e}_\alpha$ which lie in a nilpotent subalgebra of $\g_3$. Such a nilpotent subalgebra is always associated to a semi-simple element \footnote{Semi-simple means that it is in the conjugation class of an element of the Cartan subalgebra.} ${\bf h}$ of $\sl_2 \oplus \g_4$ such that
\begin{equation}  {\bf h }\,  {\bf e}_\alpha  :=  [  {\bf h } , {\bf e}_\alpha ]  = p_\alpha {\bf e}_\alpha \ , \qquad 1\le p_\alpha \le n \ , \end{equation}
where $n$ defines the maximal possible eigenvalue of $\mbox{ad}_{\bf h}$ in $\g_3$.  This implies for instance the equation
\begin{equation} \prod_{i=1}^n ( h - i ) \, P = 0 \label{FirstOrderP} \ , \end{equation}
which defines a first order constraint between the components ${\rm w}$, $Z$, $Z_a$ and $\Sigma^a$ of $P$. In order to be consistent with the equations of motion and the Bianchi identity \eqref{eom-bian-coset}, the covariant derivative of the generator ${\bf h}$ must satisfy
\begin{equation} \sum_{i=1}^n \prod_{j=i+1}^n ({\bf h} -j) d_B {\bf h}  \prod_{k=1}^{i-1} ( {\bf h} -k) \wedge P  = 0 \ , \quad  \sum_{i=1}^n \prod_{j=i+1}^n ({\bf h} -j) d_B {\bf h}  \prod_{k=1}^{i-1} ( {\bf h} -k) \star  P  = 0  \ . \end{equation}
These equations are satisfied if $d_B {\bf h}$ also lies in the nilpotent algebra defined by ${\bf h}$,  or equivalently that
\begin{equation} \prod_{i=1}^n ( \mbox{ad}_{\bf h} - i ) d_B {\bf h} = 0 \ . \label{AuxilH} \end{equation}
In general, one can always choose the generators ${\bf h}$ such that only its components $\lambda$ and $\Lambda_a$ do not vanish. Equation \eqref{AuxilH} can be viewed as first order equations for these auxiliary components, which can be solved to determine their evolution in space in terms of the physical fields. As mentioned above, equation \eqref{FirstOrderP} defines first order equations for the physical fields which contain these auxiliary components  $\lambda$ and $\Lambda_a$ and determine $dU + \frac{i}{2} e^{2U} \star d \omega$ and $e^a_i dt^i$ in terms of $e^{U} Z(\star F)$ and $e^{U} Z_a(\star F)$, plus some possible constraints on the latter if the dimension of the coset component of the nilpotent algebra defined by ${\bf h}$ is strictly less than $2n_v + 2$.

For BPS solutions one has $\Lambda_a = 0 $ and $\lambda = e^{i\alpha}$, where the phase $\alpha$ defines the covariantly constant spinors as in \cite{Denef:2000nb}, and \eqref{AuxilH} is equivalent to the equation 
\begin{equation} d \alpha  + Q - \frac{1}{2} e^{2U} \star d \omega = 0 \, .\end{equation}
The non-BPS solutions with vanishing central charge at the horizons are described in a similar fashion, with $\lambda = 0 $ and a normalised rank one $\Lambda_a$ (\ie $c^{abc} \Lambda_b \Lambda_c = 0 $ and $\Lambda_a \bar \Lambda^a = 1$) \cite{Bossard:2010mv}. We will now discuss the specific examples of the nilpotent orbits associated to the systems describing respectively composite and single centre non-BPS black holes with a non-vanishing central charge at the horizon.

\subsubsection*{Composite nilpotent elements}
The composite non-BPS solutions admit a scalar momentum $P$ which lies in the positive grade
components of the graded decomposition of the coset component $ {\bf 2} \otimes \mathds{R}^{2n_v+2}$,
associated to an element ${\bf h}$ of $\g_4$ that leads to the decomposition
\begin{equation} \g_4 \cong (\mathds{R}^{n_v} )^\ord{-2} \oplus \scal{ \gl_1 \oplus \g_5 }^\ord{0} \oplus (\mathds{R}^{n_v} )^\ord{2}\,, \end{equation}
for $\g_4$ itself and 
\begin{equation}  {\bf 2} \otimes \mathds{R}^{2n_v+2} \cong {\bf 2}^\ord{-3} \oplus ({\bf 2}\otimes\mathds{R}^{n_v} )^\ord{-1} \oplus ({\bf 2}\otimes\mathds{R}^{n_v} )^\ord{1}\oplus  {\bf
2}^\ord{3}\,, \end{equation}
for the coset component, {\it i.e.} $P \in({\bf 2}\otimes\mathds{R}^{n_v} )^\ord{1}\oplus  {\bf
2}^\ord{3}$. Such an element ${\bf h}$ can always be chosen to be Hermitian (\ie to lie in $\g_4
\ominus ( \mathfrak{u}(1) \oplus \k_4 )$) so that it is realised for
$\Lambda_a({\bf h})=\Omega_a$, where $\Omega_a$ satisfies
\begin{equation} \N[\Omega]  \bar \Omega_a = \frac{1}{2} c^{abc} \Omega_b \Omega_c \ . \qquad
\bar \Omega^a \Omega_a = 3  \ . \label{Identity} \end{equation}
Equivalently, $\Omega_a$ is in the $U(1) \times K_4 $ orbit of the Jordan algebra identity.

More explicitly, one finds the following action on the coset component 
\begin{equation} {\bf h} {\rm w} = \Omega_a \bar Z^a \quad {\bf h} Z = \Omega_a \Sigma^a
\quad {\bf h} \bar Z^a = \bar \Omega^a  {\bf w} + c^{abc} \Omega_b \bar \Sigma_c
\quad {\bf h} \Sigma^a = \bar \Omega^a  Z  + c^{abc} \Omega_b Z_c \,.\end{equation}
Considering the grade three part of $P$, from the equation $[ {\bf h} , P^\ord{3}] = 3 P^\ord{3} $ one obtains the
solution
\begin{equation} Z^\ord{3} = \N[\Omega] \bar {\rm w}^\ord{3} \qquad  \bar Z^{a\, \ord{3}} =
\bar \Omega^a {\rm w}^\ord{3} \qquad \Sigma^{a\, \ord{3}} = \N[\Omega] \bar
\Omega^a   \, \bar {\rm w}^\ord{3} \label{OrdreTrois}\,, \end{equation}
for an arbitrary ${\rm w}^\ord{3}$. Similarly, from the equation $[ {\bf h} , P^\ord{1}]
= P^\ord{1} $ for the grade one part, one obtains the solution
\begin{equation} {\rm w}^\ord{1} = \Omega_a \bar Z^{a\, \ord{1}} \qquad Z^\ord{1} = -
\N[\Omega] \bar \Omega^a Z_a^\ord{1} \qquad \Sigma^{a\, \ord{1}} =  c^{abc}
\Omega_b Z_c^\ord{1} - \N[\Omega] \bar\Omega^a \bar\Omega^b Z_b^\ord{1}\,,
\label{OrdreUn} \end{equation}
for an arbitrary $Z^\ord{1}_a$.

Considering a general linear combination of these two solutions one concludes
that $Z$ and $Z_a$ are arbitrary, whereas ${\rm w}$ and $\Sigma^a$ are
determined as
\begin{equation} {\rm w} = \frac{1}{2} \scal{ \Omega_a \bar Z^a - \N[\Omega] \bar Z }\,, \qquad
\Sigma^a = c^{abc} \Omega_b Z_c + \frac{1}{2} \bar \Omega^a \scal{ Z -
\N[\Omega] \bar \Omega^b Z_b }\,, \label{NonBPSfirstOrder} \end{equation}
which are the explicit first order relations for the scalar momenta.

These contain the auxiliary components $\N[\Omega]$ and $\Omega_a$ of ${\bf h}$ in \eqref{Identity}, which can be viewed as defining a very small vector $R$ of unit mass (\ie $I_4(R) = \partial I_4(R) = \partial^2 I_4(R)|_{\rm ad} = 0 $ and $|Z(R)| = 1$) through
\begin{equation} Z(R) = \N[\Omega] \qquad Z_a(R) = \Omega_a\,. \label{R-def} \end{equation}
The flow equations for these fields are given by \eqref{AuxilH}, which in this system reduces to
\begin{equation} [ {\bf h} , d_B {\bf h} ] = 2\, d_B {\bf h} \ . \label{GradeDeux}  \end{equation}
Using the explicit form of $B$ \eqref{Bsl2}-\eqref{Bk4} and the first order constraint \eqref{NonBPSfirstOrder}, one computes the components of $d_B {\bf h}$ as
\begin{eqnarray} \gamma(d_B {\bf h}) &=&  \frac{2}{3} \, \Im[\bar\Omega^a Z_a]\,,
\CR
\Lambda_a(d_B {\bf h})  &=& Z_a(dR) + \N[\Omega] e_{a\bar \imath} d\bar t^{\, \bar \imath} + c_{abc} \bar \Omega^b e^c_i dt^i - \frac{i}{2} e^{2U} \star d\omega \Omega_a \CR
&=&  Z_a( d R) + \mbox{Re}[ \N[\bar \Omega] \Omega_b e^b_i dt^i ] Z_a(R) - \Scal{ Z_a + \N[\Omega]
 c_{abc}\bar \Omega^b \bar Z^c
- \Omega_a\,\Omega_b \bar Z^b}
\CR
G^a{}_b(d_B{\bf h})  &=& c^{ace} c_{bde} \scal{ \bar \Omega^d Z_c - \Omega_c \bar Z^d } + \Omega_b \bar Z^a - \bar \Omega^a Z_b - \frac{2i}{3} \delta^a_b \mbox{Im}[ \bar \Omega^c Z_c ] \label{dh-compon}
\, ,
\end{eqnarray}
where we explicitly separated the terms depending on the derivative of the vector $R$. It is now straightforward (though cumbersome) to compare \eqref{GradeDeux} with the above relations, using that
\begin{eqnarray} && c^{ace} c_{bde} \bar \Omega^d \scal{ \N[\Omega] c_{cfg} \bar \Omega^f \bar Z^g - \Omega_c \Omega_f \bar Z^f } -  \bar \Omega^a \scal{ \N[\Omega] c_{bcd} \bar \Omega^c \bar Z^d - \Omega_b \Omega_c \bar Z^c }  \CR
&=&- c^{ace} c_{bde} \Omega_c \bar Z^d + \Omega_b \bar Z^a \,, \end{eqnarray}
which follows by \eqref{symmetric}. The result is that \eqref{GradeDeux} is satisfied provided that
\begin{equation} d R = -  \mbox{Re}[ \N[\bar \Omega] \Omega_b e^b_i dt^i ] R \,,  \label{eq:R-der} \end{equation}
which implies that there exist a constant symplectic vector $\hat{R}$ such that
\begin{equation} R = \frac{\hat{R}}{|Z(\hat{R})|} \label{eq:R-rescale} \ . \end{equation}
We conclude that the generator ${\bf h}$ is in this case determined by a constant very small projective vector $\hat{R}$ and the scalar fields such that
\begin{equation} \Lambda_a({\bf h}) = \frac{ Z_a(\hat{R})}{|Z(\hat{R})|} \, . \label{OmegaRhat}  \end{equation}
One can now return to \eqref{NonBPSfirstOrder}, which becomes a first order flow equation for the scalars $e^U$, $\star d\omega$ and $dt^i$ in terms of the gauge fields and the constant vector $\hat R$.

\subsubsection*{Single centre nilpotent elements}
The composite non-BPS system above can be consistently reduced to a system describing the single centre class of solutions. The associated graded decomposition consists in
breaking furthermore the $\sl_2$ algebra by introducing a non-compact generator
${\bf h}_*$ defined such that its only nonvanishing component is $\lambda = e^{i \alpha}$. The action of this generator follows from \eqref{sl2-alg} as
\begin{equation} {\bf h}_* {\rm w} = e^{-i \alpha} Z\,, \quad 
{\bf h}_* Z =  e^{i \alpha}  {\rm w}\,, \qquad 
{\bf h}_* \bar Z^a = e^{-i \alpha}  \Sigma^a\,, \quad 
{\bf h}_* \Sigma^a = e^{i \alpha} \bar Z^a\,, \end{equation}
and, by its own, it would define the BPS system. The single-centre non-BPS solution is defined in the
positive grade component of the graded decomposition associated to the generator
$\frac{1}{2} ( {\bf h} + {\bf h}_*)$
\begin{equation}  {\bf 2} \times \mathds{R}^{2n_v+2}  \cong \mathds{R}^\ord{-2} \oplus (\mathds{R} \oplus \mathds{R}^{n_v} )^\ord{-1} \oplus \scal{\mathds{R}^{n_v}  \oplus \mathds{R}^{n_v} }^\ord{0} \oplus
(\mathds{R}\oplus \mathds{R}^{n_v} )^\ord{1}\oplus  \mathds{R}^\ord{2}\,,
\label{SingleCentreGD} \end{equation}
where ${\bf h}$ is the generator that defines the composite system above.
One can straightforwardly compute that the solution (\ref{OrdreTrois}) decomposes
into $[ {\bf h}_* , P^\ord{3}_\pm ] = \pm P^\ord{3}_\pm$ according to
\begin{equation} {\rm w}^\ord{3}_\pm = \pm e^{-i \alpha} \N[\Omega] \bar {\rm w}^\ord{3}_\pm
\end{equation}
and these two solutions define the grade $1$ and the grade $2$ singlets in
(\ref{SingleCentreGD}). On the other hand, the grade $1$ component of
(\ref{SingleCentreGD}) in $\mathds{R}^{n_v}$ corresponds to the solution
(\ref{OrdreUn}) satisfying moreover $[ {\bf h}_* , P^\ord{1}_+ ] = P^\ord{1}_+ $, or explicitly
\begin{equation} \bar Z^{a\, \ord{1}}_+ = e^{-i \alpha} \scal{ c^{abc} \Omega_b Z_{c\,
+}^\ord{1} - \N[\Omega] \bar\Omega^a \bar\Omega^b Z_{b\, +}^\ord{1} }\,. \end{equation}
Summing up the two solutions, one obtain that the single centre non-BPS momenta
satisfy (\ref{NonBPSfirstOrder}) for $Z$ and $Z_a$ constrained to satisfy the
phase dependent equation
\begin{equation} \bar Z^a - \N[\Omega] \bar \Omega^a \bar Z = e^{-i \alpha} \scal{ c^{abc}
\Omega_b Z_c + \bar \Omega^a ( Z - \N[\Omega]  \bar \Omega^b Z_b )}
\label{ZAconstraint} \,, \end{equation}
which represents a constraint on the physical degrees of freedom that is necessary to reduce to single centre solutions.

This is expected for a single centre solution, since the element
$\Omega_a$ is defined by its overall phase and the angle $K_4 /K_5$ ( $K_5$ being  the maximal compact subgroup of $G_5$, and therefore the stabilizer of $\Omega_a$ in $K_5$), that is $n_v$ real
parameters in total. The constraint (\ref{ZAconstraint}) defines precisely $n_v$
real equations, such that one can think of it as determining $\Omega_a$ in terms
of the central charge $Z$, its derivatives and the `BPS phase' $\alpha$.

On the other hand, when viewed as a constraint on the charge vector it is simple to see that \eqref{ZAconstraint}
reduces its components by half. Defining the combination
$K_a=Z_a - \N[\bar\Omega] \Omega_a Z$, the constraint becomes
\begin{equation} \label{reduced-constraint}
\bar K^a
= e^{-i \alpha} \scal{ c^{abc}
\Omega_b K_c
- \N[\Omega]\, \bar \Omega^a\, \bar \Omega^b K_b }
\equiv \iota (  K_a) \,.
\end{equation}
In the right-hand-side we defined the operation $\iota$, which is an anti-involution
\begin{eqnarray}
\iota (\iota  (\bar K^a ))
&=& e^{-i \alpha} \scal{ c^{abc}
\Omega_b \iota (\bar K^c)
- \N[\Omega]\, \bar \Omega^a\, \bar \Omega^b \iota( \bar K^b)  }
\CR
&=& c^{abc}\Omega_b c_{cde} \bar \Omega^d \bar K^e
- \bar \Omega^a\, \Omega_b \bar K^b
\CR
&=& \bar K^a
\,,
\end{eqnarray}
where we used \eqref{Identity} in the last equation. In what follows we will elaborate on these points of view of the constraint \eqref{ZAconstraint}, in connection to the various aspects of the solutions.

We close this section by giving the analog of the consistency condition
\eqref{GradeDeux} for the generator ${\bf h}_*$. In this case, only the $\sl_2$ components are important, and using \eqref{Bsl2} one computes that
\begin{equation} \rho(d_B {\bf h}_*) = - 2 \, \mbox{Im}[ e^{-i \alpha} Z]  \ , \qquad \lambda(d_B {\bf h}_*) = D  e^{i \alpha}
 +\tfrac{i}{2}\,\mathrm{e}^{2U} \star d \omega \, e^{i \alpha}
\,,
\label{dh-star-compon}
\end{equation}
where the covariant derivative $D$ is the K\"ahler covariant derivative in
\eqref{Kah-metr-der}, consistent with the unit K\"ahler weight of the phase
$e^{i \alpha}$. Imposing that this is a grade two element of the $\sl_2$
algebra, \ie that
\begin{eqnarray}
2\,\rho( d_B {\bf h}_*)&=&
i (e^{i \alpha} \bar\lambda( d_B {\bf h}_*) -e^{-i \alpha} \lambda( d_B {\bf h}_*))
\,,
\qquad
2\,\lambda( d_B {\bf h}_*) = 2\,i\,e^{i \alpha} \rho( d_B {\bf h}_*)
\,,
\end{eqnarray}
turns out to be equivalent to the purely imaginary condition
\begin{equation}
e^{-i \alpha}\,D ( e^{i \alpha} )
 +\tfrac{i}{2}\,\mathrm{e}^{2U} \star d \omega
= -2\,i\,\Im[e^{-i  \alpha} Z]\,,
\end{equation}
which fixes the phase $\alpha$ in terms of the physical degrees of freedom
through
\begin{equation} \label{alpha-fix1}
d\alpha+Q  +\tfrac{1}{2}\mathrm{e}^{2U} \star d \omega
= -2\,\Im[e^{-i \alpha} Z] \,.
\end{equation}
Since we also have the phase $\N[\Omega]$, it is natural to define the K\"ahler invariant phase $e^{i  \alpha} \N[\bar \Omega]$. From \eqref{OmegaRhat} we find that
\begin{equation} d \, \mbox{arg}[ \N[\bar \Omega] ] - Q + \mbox{Im}[ \N[\bar \Omega] \Omega_a e^a_i dt^i ] = 0 \ , \end{equation}
and using moreover \eqref{NonBPSfirstOrder} one obtains that  
\begin{multline}  d \scal{ \alpha +  \mbox{arg}[ \N[\bar \Omega] ] } + \frac{1}{2} e^{2U} \star d \omega \\ + \mbox{Im}\bigl[ \N[\bar \Omega] \Omega_a e^a_i dt^i + e^{-i  \alpha} \N[\Omega] \scal{ dU - \tfrac{i}{2} e^{2U} \star d \omega - \N[\bar \Omega] \Omega_a e^a_i dt^i } \bigr] = 0 \ . \end{multline}
Through \eqref{NonBPSfirstOrder}, \eqref{ZAconstraint} also implies a reality constraint on the scalar field momenta
\begin{multline} \N[\bar \Omega]  e^a_i dt^i + e^{i  \alpha}  \N[\bar \Omega]  \scal{ - c^{abc} \Omega_b e_{c \bar \imath} d \bar t^{\, \bar \imath} + \N[\Omega] \bar \Omega^a \bar \Omega^b e_{b \bar \imath} d \bar t^{\, \bar \imath} } \\ = \bar \Omega^a \Scal{ dU - \tfrac{i}{2} e^{2U} \star d\omega + e^{i  \alpha}  \N[\bar \Omega]  \scal{  dU + \tfrac{i}{2} e^{2U} \star d\omega }}  \label{RealT} \end{multline}
and in particular
\begin{equation} \N[\bar \Omega] \Omega_a e^a_i dt^i + e^{i  \alpha} \bar \Omega^a e_{a \bar \imath} d \bar t^{\, \bar \imath} = 3 \Scal{ dU - \tfrac{i}{2} e^{2U} \star d \omega + e^{i  \alpha} \N[\bar \Omega] \scal{  dU + \tfrac{i}{2} e^{2U} \star d \omega}} \ . \label{RealTtrace}  \end{equation}
Therefore one obtains finally
\begin{eqnarray} \hspace{-10mm} d \scal{ \alpha +  \mbox{arg}[ \N[\bar \Omega] ] }  &=& e^{2U} \star d \omega - 2 \mbox{Im}\bigl[ e^{i  \alpha} \N[\bar \Omega] \scal{ dU + \tfrac{i}{2} e^{2U} \star d \omega }\bigr] \CR
&=& \Scal{ 1- \cos\scal{ \alpha +  \mbox{arg}[ \N[\bar \Omega] ] } } \left( e^{2U}\star d \omega - \frac{2}{ \tan\scal{ \frac{\alpha +  {\scriptstyle \rm arg}[ \N[\bar \Omega] ] }{2}}} d U \right)\,,  \end{eqnarray}
which can be integrated to
\begin{equation} \star d \omega = d \frac{-e^{-2U} }{\tan\scal{ \frac{\alpha +  {\scriptstyle \rm arg}[ \N[\bar \Omega] ] }{2}}}   \equiv d M \ .  \label{omegaM} \end{equation}
We conclude that the angular momentum is given in terms of a harmonic function $M$ dual to the Kaluza--Klein vector, which is a known characteristic feature of single centre solutions \cite{Bena:2009ev}. The phase $e^{i  \alpha} \N[\bar \Omega]$ is determined in this way in terms of spacetime fields as
\begin{equation} e^{i  \alpha} \N[\bar \Omega] = \frac{( - M + i e^{-2U})^2}{e^{-4U} + M^2} \ .  \label{alphaM} \end{equation}

\section{Duality covariant form of the non-BPS black hole solution}
\label{sec:single-class}

In this section we solve explicitly the flow equations in the first order system describing non-interacting non-BPS black holes discussed in the last section. We will see that in this case the vector fields $w$ and $\omega$ carrying the electromagnetic charges and the angular momentum are simply sourced by harmonic functions, although the vector fields satisfy a quadratic constraint such that they only depend on $n_v+1$ harmonic functions, instead of $2n_v + 2$ in the BPS system \cite{Denef:2000nb}. Incidentally we exhibit that this first order system reduces to a linear system of differential equations. This is a necessary step towards the explicit solution of the non-BPS composite system describing interacting non-BPS black holes. After presenting the procedure
of integrating \eqref{NonBPSfirstOrder} combined with the reality constraint 
\eqref{ZAconstraint}, we briefly discuss the physical properties of the solutions.

\subsection{Integrating the first order equations}
The starting point is the solution of the nilpotency condition
\eqref{NonBPSfirstOrder}, written explicitly as a first
order  system for the scalars and the metric degrees of freedom
\begin{eqnarray} dU + \frac{i}{2} e^{2U} \star d \omega &=& - \frac{1}{2} e^U \scal{
\Omega_a
\bar Z^a(\star F) - \N[\Omega] \bar Z(\star F)} \label{Grav} \ , 
\CR
- e^a_i d t^i &=& c^{abc} \Omega_b  e^U Z_c(\star F) + \tfrac12 \bar \Omega^a
e^U
\scal{ Z(\star F) - \N[\Omega] \bar \Omega^b  Z_b(\star F) }\,, \label{ScalLin}
\end{eqnarray}
where $F$ is the spatial component of the field strengths defined in \eqref{gauge-decop}. For later reference, we give the inverse
relations for the field strengths
\begin{eqnarray} e^U Z(\star F) &=& \frac{1}{2} \N[\Omega] \Scal{ dU - \frac{i}{2} e^{2U}
\star d \omega } - \frac{1}{2} \Omega_i d t^i \label{ZstarF} \ , \CR
e^U Z_a(\star F) &=& -c_{abc} \bar \Omega^b e^c_i d t^i
+\frac{1}{2} \N[\bar \Omega] \Omega_a \Omega_i dt^i  - \frac{1}{2}
\Omega_a \Scal{  dU - \frac{i}{2} e^{2U} \star d \omega } \ , \end{eqnarray}
where we used the short-hand notation $\Omega_i = e_i^a \Omega_a$.

\subsubsection*{Electromagnetic scalar potentials}
In order to solve this system, we combine the information on the derivative of
$R$ given by \eqref{eq:R-der} with \eqref{ZstarF} to construct the gauge field momenta as
\begin{eqnarray}
d\zeta&\equiv& 2\,e^{2U}\Re[\bar Z(\star F) \cV + \bar Z^i(\star F) D_{i}
\cV] \CR
&=& d\,\scal{ e^{U} \Re [ \N[\bar \Omega]  \cV
 -\bar \Omega^i  D_i \cV]}
+\tfrac{1}{2}e^{U}\,\Im[ \N[\bar \Omega] \Omega_i d t^i] \, R
+\tfrac{1}{4} e^{3U} \star d \omega\,R
\label{eq:dzeta}\,,
\end{eqnarray}
where we made extensive use of the special geometry identities \eqref{SpecialGeoId}. The first term is manifestly a total derivative,
whereas the others are along the very small vector $R$, and must therefore combine into the derivative of a single function. This implies the existence of a function $M$ such that
\begin{equation}
e^{U}\,\Im[\N[\bar\Omega]\, \Omega_i d t^i]
+\tfrac{1}{2} e^{3U} \star d \omega
= M\,e^{3U}\,\Re[ \N[\bar\Omega] \Omega_i d t^i]
- d (M\,e^{3U}) \,.
\label{ImOdt}
\end{equation}
And indeed, in the `single centre' system one shows using \eqref{alphaM} in \eqref{RealTtrace} that $M$ is the function that determines the angular momentum in \eqref{omegaM}. It follows that the gauge field momenta take the form
\begin{equation} \zeta =  e^U \Re [ \N[\bar \Omega]  \cV
 -\bar \Omega^i  D_i \cV]
- \frac{1}{2} e^{3U} M R\,, \label{Zetas} \end{equation}
with the corresponding central charges given by
\begin{equation} Z( \zeta) =
\frac{i}{2}\,e^U\, ( 1 + i\,e^{2U} M) \N[\Omega]\,,
 \qquad Z_a( \zeta) = \frac{i}{2}\,e^U\, ( 1 + i\,e^{2U} M) \Omega_a\,,
\label{Zzeta}\end{equation}
for later reference.

Using \eqref{Grav} and the structure of the \eqref{Zetas}, one shows that one vector is always
trivial, because 
\begin{eqnarray} \Iprod{\hat R}{\zeta}
&=& -2 e^U |Z(\hat{R})|\,,\end{eqnarray}
whereas, taking the imaginary part of (\ref{Grav}) one gets
\begin{eqnarray} e^{2U} \star d\omega
&=& -\frac{e^U}{2 |Z(\hat{R})|}\, \Iprod{\hat R}{\star F}\,, \end{eqnarray}
and using (\ref{gauge-decop}), one gets therefore that
\begin{equation} \Iprod{\hat R}{dw}= 0\,.
\label{Rdw}
\end{equation}
Note that this property does not require the reality constraint \eqref{ZAconstraint} and is also valid for composite non-BPS solutions \cite{Bossard:2012ge}.
\subsubsection*{The linear system}
One can now combine \eqref{Zzeta} and \eqref{Rdw} to disentangle the term
proportional to $\star d\omega$ in the definition of the scalars, such that
(\ref{ScalLin}) becomes
\begin{eqnarray}
-e^a_i d t^i &=& c^{abc} \Omega_b  e^U Z_c(\star dw) + \tfrac12 \bar \Omega^a e^U
\scal{ Z(\star dw) - \N[\Omega] \bar \Omega^b  Z_b(\star dw) }  \CR
&& \hspace{55mm} + \tfrac12  e^{4U} (- M + i e^{-2U} ) \N[\Omega] \bar \Omega^a
 \star d \omega\,,
\end{eqnarray}
which, using the constraint \eqref{ZAconstraint}, can be rewritten as
\begin{eqnarray}\label{eq:nbps-scals}
-e^a_i d t^i &=&  e^U e^{i \alpha} (\bar Z^a(\star dw) - \N[\Omega] \bar
\Omega^a \bar Z(\star dw) )
 -\N[\Omega]\, \bar \Omega^a\, \Scal{ dU - \frac{i}{2} e^{2U} \star d \omega }
\,.
\end{eqnarray}

Applying the same procedure on \eqref{ZstarF}, one obtains the the inverse relations of \eqref{eq:nbps-scals} for the central charges $Z(dw)$ and $Z_a(dw)$. The charge vectors $dw$ can then be straightforwardly constructed with
the result
\begin{eqnarray}
dw&\equiv& 2\,\Im[-\bar Z(dw)\cV + \bar Z^a(dw)D_a\cV] \CR
&=&-2\,e^{-U} \Im[e^{-i \alpha} (-\star dU +\tfrac{i}2e^{2U}d\omega)\cV
   + e^{-i \alpha} d t^iD_i\cV  +\tfrac12 \, \mu \,\cR] \,,
\end{eqnarray}
where we used the shorthands
\begin{eqnarray}
 \cR &=& -\N[\bar\Omega] \cV +\bar\Omega^{i}D_i\cV\,,
\CR
\mu 
 &=& - \frac{1}{2} e^{4U} d ( e^{-4U} + M^2 ) + e^{-i  \alpha} \N[\Omega] \scal{ \star dU - \tfrac{i}{2} e^{2U} d\omega - \N[\bar \Omega] \Omega_i \star d t^i } \,.
\end{eqnarray}

At this stage we have exhausted the constraints implied by the existence of the constant $\hat R$, and it is important to find a constant vector Darboux conjugate to $\hat{R}$ in order to be able to decompose conveniently $dw$. This is indeed possible, using equations \eqref{eq:R-der}, \eqref{RealT}, \eqref{omegaM}, \eqref{alphaM} and \eqref{ImOdt}, which allow one to show that 
\begin{align}   \hat{R}^* =
 |Z(\hat{R})|^{-1}&\,
\mbox{Re}\Bigl[
\bar Y^3\,\N[\bar\Omega]\,{\mathcal V}
+|Y|^2\bar Y\bar \Omega^{i} D_i \cV \Bigr]\,, 
\label{Rstar} \\
Y \equiv &\, ( 1 + i\,e^{2U} M) \label{Y-def}\,,
\end{align}
is constant, in the following way
\begin{eqnarray} d \hat{R}^* &=& - 2 \hat{R}^*\,  \mbox{Re}[ \N[\bar \Omega] \Omega_i d t^i ] + \frac{1}{|Z(\hat{R})|} \mbox{Im}\bigl[ 3 \bar Y^2 \N[\bar\Omega] \cV + \bar Y ( 2 Y - \bar Y ) \bar \Omega^i D_i \cV \bigr] d ( e^{2U} M ) \CR
&& \hspace{20mm}  + \frac{2}{|Z(\hat{R})|} \mbox{Re}\bigl[ \bar Y^2 \bar \Omega_{\bar \imath} d \bar t^{\, \bar \imath} \cV + \bar Y \scal{ \bar Y \N[\bar \Omega] d t^i + Y e^i_a c^{abc} \Omega_b e_{c\bar \jmath} d \bar t^{\, \bar \jmath}} D_i \cV \bigr] \CR
&=& - 2 \hat{R}^* \,  \mbox{Re}[ \N[\bar \Omega] \Omega_i d t^i ] + \frac{3}{ |Z(\hat{R})|} \mbox{Im}\bigl[ \bar Y^2  \N[\bar\Omega] \cV + |Y|^2   \bar \Omega^i D_i \cV \bigr] d ( e^{2U} M ) \CR
&& \hspace{20mm}  + \frac{2}{|Z(\hat{R})|} \mbox{Re}\bigl[ \scal{ \bar Y^2 \N[\bar \Omega] \cV + |Y|^2 \bar \Omega^i D_i \cV  } \scal{ \N[\bar \Omega] \Omega_i dt^i } \bigr] \CR
&=& 0 \,. \end{eqnarray}
This vector is indeed mutually non-local with $\hat R$, since their inner product, $\Iprod{\hat{R}^*}{\hat R}=-4$ and is also very small, \ie its central charges satisfy \eqref{Identity}, as a consequence of its definition.

One can now project $dw$ along this new vector to find
\begin{eqnarray}
\Iprod{\hat{R}^*}{dw}
&=& -\star d(e^{-U}|Z(\hat R)|^{-1}\,|Y|^2) \equiv -\star dV\,,
\label{V-def}
\end{eqnarray}
where we defined the distinguished harmonic function $V$, whose pole will carry the linear combination $\Iprod{\hat{R}^*}{\Gamma}$ of the physical charges. Note that $V$ can equivalently be defined as
\begin{equation} V  \equiv e^{-U} \sqrt[3]{ \frac{ 4 |Z(\hat{R}^*)|^2}{|Z(\hat{R})|}} \ , \end{equation} 
and because the central charge of a very small vector is nowhere vanishing in moduli space, it follows that for a regular extremal solution, the function $V$ is strictly positive and $\Iprod{\hat{R}^*}{\Gamma} < 0 $. Combining this with \eqref{eq:R-der}, \eqref{ImOdt} and \eqref{eq:nbps-scals} we
can determine the combination $\Omega_i d t^i$ in terms of $V$, the metric components and the phase $e^{i\alpha}$ as
\begin{eqnarray}
\Re[ \N[\bar\Omega]\, \Omega_i \,d t^i]
&=&-V^{-1} dV -\, dU +|Y|^{-2}\,d |Y|^{2}
 \,,
\CR
e^{-i \alpha} \Omega_i \,d t^i
&=& -\bar Y\,\Re[ \N[\bar\Omega]\, \Omega_i \, d t^i]
-\tfrac32\, e^{-i \alpha}\,\N[\Omega]\, d Y
 \,.\label{eq:Omega-dt}
\end{eqnarray}

We can now use this information to further simplify the expression for $dw$ by
adding multiples of the two constant vectors with appropriate coefficients. It turns out that the most
suggestive form is obtained by subtracting a multiple of $\hat R$ and arranging that the sign of the $\hat R$-component
is flipped. This clearly reflects the situation one encounters in the known explicit solution defined through the almost-BPS system \cite{Khuri:1995xq,Bena:2009ev}, and indeed one finds that
\begin{eqnarray}
dw -  2\,\tfrac{\Iprod{\hat{R}^*}{dw}}{\Iprod{\hat{R}^*}{\hat R}}\,\hat R
&=&
-2\,\Im[e^{-U} e^{-i \alpha} \left(
 (-\star dU +\tfrac{i}2e^{2U}d\omega)\cV + d t^iD_i\cV\right)]
\CR
&&
+2\, V\,e^{U}\,
\Im\left( i\,e^{-i \alpha} \cV\right)\,\star d(\tfrac{M}{V})
+ \star d\left(\tfrac{M}{V}\right)\, \hat{R}^*
\CR
&=&
-2\,\Im\star\tilde{D}(e^{-U} e^{-i \alpha} \cV)
+ \star d\left(\tfrac{M}{V}\right)\, \hat{R}^*
\label{eq:dw-rewrite} \,,
\end{eqnarray}
where we defined the modified covariant derivative
\begin{equation}
\tilde{D}(e^{-U} e^{-i \alpha} \cV)
=\left[ d + i\,( Q + d\alpha +\tfrac{1}2e^{2U}\star d\omega
          - \,e^{2U}\, d(\tfrac{M}{V})\, V ) \right]
(e^{-U} e^{-i \alpha} \cV)\,.
\end{equation}
The form of this covariant derivative is exactly such that the corresponding composite connection is trivial by use of 
\eqref{alpha-fix1}, \eqref{ZstarF} and \eqref{eq:Omega-dt}
\begin{equation} \label{alpha-fix2}
Q + d\alpha +\tfrac{1}2e^{2U}\star d\omega
          -\,e^{2U}\, d(\tfrac{M}{V})\, V =0\, .
\end{equation}
It then follows that \eqref{eq:dw-rewrite} takes the form
\begin{equation} dw -  2\,\tfrac{\Iprod{\hat{R}^*}{dw}}{\Iprod{\hat{R}^*}{\hat R}}\,\hat R
=-\star d \left[2\,\Im \left(e^{-U} e^{-i \alpha} \cV\right)
- \left(\tfrac{M}{V}\right)\, \hat{R}^*\right]\,,\label{dw-fin}
\end{equation}
which implies that the vector fields are defined in terms of harmonic functions as
\begin{equation} dw=\star d\cH\, ,\label{harmonic-def} \end{equation}
such that for instance $\langle \hat{R}^* , \cH \rangle = - V $. Using this back in \eqref{dw-fin} one finds that the scalars are given by
\begin{equation}
2\,\Im(e^{-U} e^{-i \alpha} \cV)=
-\cH + 2\,\tfrac{\Iprod{\hat{R}^*}{\cH}}{\Iprod{\hat{R}^*}{\hat R}}\,\hat R
 - \tfrac{M}{\Iprod{\hat{R}^*}{\cH}}\, \hat{R}^* \,.
\label{scal-sol}
\end{equation}
Note that this result only depends on harmonic functions, where the angular momentum harmonic function only appears through its ratio with $V$, as in \cite{Galli:2010mg}, and the harmonic functions $\cH$ control the electric and magnetic charges. However these harmonic functions are not all independent as in the BPS system, but are subject to algebraic constraints that reduce them to $n_v+1$ independent functions, as we show in the next paragraph using the constraint \eqref{ZAconstraint}. This constraint is rather non-linear because of its dependence on the scalar fields, and we will now rewrite it as a quadratic constraint in the harmonic functions themselves.  
\subsubsection*{Quadratic constraint}
For this purpose will make use of the quartic
invariant of $\N=2$ supergravity coupled to a symmetric scalar manifold, which reads \cite{Ferrara:2006yb}
\begin{eqnarray}
I_4(\Gamma)&=& \frac{1}{4!} t^{MNPQ}\Gamma_M\Gamma_N\Gamma_P\Gamma_Q
\nonumber\\
         &=& - (p^I\,q_I)^2 + 4\,q_0\,\N[p] - 4\,p^0\,\N[q] 
             + c_{ijk}p^jp^k\,c^{ilm}q_lq_m\,, \label{I4-ch}
\end{eqnarray}
in terms of the cubic norm. Here, $t^{MNPQ}$ is a completely symmetric tensor,
and $M,N,\dots$ are symplectic indices that encompass both the upper
and the lower components in \eqref{eq:dual-gauge}. The absolute value of this
expression is known to determine the entropy of static black holes for any value of
the charges. We also define the lift of symplectic indices through the symplectic form
\begin{equation} \Gamma_M = \left(\begin{array}{c} p^I\\q_I \end{array}\right) \; , \qquad \Gamma^M = \left(\begin{array}{c} -q_I \\ p^I  \end{array}\right) \; ,\end{equation}
such that 
\begin{equation} \Iprod{\Gamma_1}{ \Gamma_2 } = \Gamma_{1 M} \Gamma_2^M =  -  \Gamma_{2 M} \Gamma_1^M \; . \end{equation}

We know already from \eqref{Rdw} that the harmonic functions $d \cH = \star d w$ must be symplectic normal to the vector $\hat R$, and because the integration constant of $\cH$ along $\hat{R}^*$ can always be reabsorbed in a redefinition of the integration constant of the function $M$ in \eqref{scal-sol}, we can assume without loss of generality that 
\begin{equation} \Iprod{\cH}{\hat R}=0\,. \label{RHzero} \end{equation}
To rewrite the constraint \eqref{ZAconstraint} we will consider the vector 
\begin{equation}  \label{I4-der}
 I^{'\,M}_{4}(\cH,\hat{R}^*) \equiv \frac{\partial^2 I_4(\cH)}{\partial \cH_M \partial \cH_N} \hat{R}^*_N = \frac{1}{2} t^{MNPQ} \cH_N \cH_P \hat{R}^*_Q \,,
\end{equation}
where $\hat{R}^*$ is the small vector defined in \eqref{Rstar} as the magnetic dual to $\hat R$. To compute the decomposition of this vector in terms of its components linear in $\cH^M$ itself and the small vectors $\hat{R}^M$ and $\hat{R}^{*M}$, it will be useful to observe the following consequence of symplectic invariance
\begin{equation} \label{I4-cmplx} \frac{\partial I_4}{\partial \bar Z}(q,p) = - i Z\scal{ \tfrac{\partial I_4}{\partial p} , -  \tfrac{\partial I_4}{\partial q}} \; , \qquad 
\frac{\partial I_4}{\partial \bar Z^a } = i Z_a\scal{ \tfrac{\partial I_4}{\partial p} , -  \tfrac{\partial I_4}{\partial q}} \; , \end{equation} 
which will allow us to compute \eqref{I4-der} in the complex basis starting from the
alternative expression for the quartic invariant of a vector in terms of its
central charges
\begin{equation} I_4 = \scal{ Z \bar Z - Z_a \bar Z^a }^2 - c_{eab} \bar Z^a \bar Z^b c^{ecd}
Z_c Z_d +  4 \bar Z \N[ Z] + 4 Z\N [ \bar Z]\,. \label{I4-def}\end{equation}
Note that, while the scalars appear explicitly in all terms, the complete
expression can be shown to be scalar independent and equal to \eqref{I4-ch}.
Using the above properties and the reality constraint \eqref{ZAconstraint} on $d\cH  = \star dw$, one can compute that
\begin{equation}  t^{MNPQ}\,\partial_\mu \cH_N\,\partial_\nu \cH_P\,\hat{R}^*_Q 
= 2\,\Iprod{\hat{R}^*}{\partial_{(\mu} \cH}\,\partial_{\nu)} \cH^M
- 4 \frac{\Iprod{\hat{R}^*}{\partial_\mu \cH} \Iprod{\hat{R}^*}{\partial_\nu \cH}}{\Iprod{\hat{R}^*}{R}} R^M \ . \label{sympl-constr-pr} \end{equation}
The interested reader can find an outline of the derivation in appendix
\ref{app-constraint}. This constraint can be integrated to the same constraint on the harmonic functions themselves, up to possible constants which do not a priori need to satisfy the constraint. Nevertheless, using \eqref{scal-sol}, and substituting the expressions \eqref{alphaM}, \eqref{Rstar}, \eqref{Y-def} and \eqref{V-def} one computes that
\begin{eqnarray} Z\scal{  \cH - \tfrac{1}{2} V \hat{R} } &=& e^{-U+i  \alpha } + \frac{M}{V} Z(\hat{R}^*) = - e^{-U} \frac{Y}{\bar Y} \N[\Omega]+ \frac{i}{2} M e^U \frac{Y^2}{\bar Y} \N[\Omega]\,,  \CR
Z_a\scal{  \cH - \tfrac{1}{2} V \hat{R} } &=&  \frac{M}{V} Z_a(R^*) = - \frac{i}{2} M e^{U} Y \Omega_a \; .  \label{HaCentral} \end{eqnarray} 
It follows that $\cH- \tfrac{1}{2} V \hat{R}$ satisfies the constraint (\ref{ZAconstraint}), which implies that $\cH$ does as well. Since $\cH$ must satisfy (\ref{ZAconstraint}), it follows that the integration constants in $\cH$ also satisfy \eqref{sympl-constr-pr} and one obtains that
\begin{equation} \frac12\, I^{'\,M}_{4}(\cH,\hat{R}^*) = \Iprod{\hat{R}^*}{\cH}\, \cH^M
- 2 \frac{\Iprod{\hat{R}^*}{\cH}^2 }{\Iprod{\hat{R}^*}{R}} R^M \, . \label{sympl-constr}\end{equation}
We can now use the fact that both $\cH$ and its derivative satisfy the above constraint, to show that they must necessarily lie in a Lagrangian subspace. Indeed, for any two vectors $\Gamma_{1,2}$ satisfying the constraint \eqref{ZAconstraint}, one can show that their inner product \eqref{inter-prod-Z}  in the complex basis can be written as
\begin{equation}
 \Iprod{\Gamma_{1}}{\Gamma_{2}}=
2\,\mbox{Im}\left[ - Z(\Gamma_{1}) \bar Z(\Gamma_{2})
  +\tfrac13\,\bar\Omega^a Z_a(\Gamma_{1})\, \Omega_b\bar Z^b(\Gamma_{2}) \right]\,,
\end{equation}
which vanishes upon requiring that $\Gamma_{1,2}$ are mutually local with $R$, as expressed by \eqref{OmegaZ-inn}. It then follows that
\begin{equation}  \Iprod{ \cH}{ d  \cH} = 0 \ . \end{equation}
We conclude that the poles of $\cH$ must be mutually local charges, and such solution cannot describe interacting black holes. Indeed, in Appendix  \ref{app-seed-sol} we show an example of this property in a specific duality frame to obtain the the $n_v+1$ harmonic functions parametrising the relevant Lagrangian subspace.

\subsubsection*{Scaling factor and moduli}
This concludes our analysis. For any very small vector $\hat{R}$ one can construct explicit solutions, after first using the asymptotic moduli in order to determine $\hat{R}^*$ from \eqref{Rstar}, which by definition will be constant. One must then solve the algebraic equation \eqref{sympl-constr}, which determines the allowed harmonic functions $\cH$. The scalars and the metric scale factor can be obtained by solving \eqref{scal-sol} in the standard way \cite{Bates:2003vx}. For instance 
\begin{equation} e^{-4U}= I_4(\cH - \tfrac{1}{2} V\,\hat R - \tfrac{M}{V}\hat{R}^*) \  . \end{equation}
Using the property that $\hat{R}^*$ is very small, it follows that the terms of order three and four in $M$ vanish and the term of order two simplifies according to \cite{Ferrara:2010ug} 
\begin{equation} \frac{1}{2} t^{MNPQ} \hat{R}^*_M \hat{R}^*_N \Gamma_P \Gamma_Q = -  \Iprod{\hat{R}^*}{\Gamma}^2 \ . \end{equation}
Considering moreover that  \eqref{sympl-constr} is satisfied for $\cH - \tfrac12\,V \hat{R}$ and contracting this constraint with $\cH - \tfrac12\,V \hat{R}$,  one obtains that the term of order one in $\hat{R}^*$ vanishes as well. Therefore 
\begin{equation} e^{-4U} = I_4(\cH  - \tfrac{1}{2} V\,\hat R)  - \tfrac{M^2}{V^2}\,\Iprod{\hat{R}^*}{\cH  - \tfrac{1}{2} V\,\hat R}^2 = I_4(\cH  - \tfrac{1}{2} V\,\hat R)   - M^2 \ . \end{equation}
Using again \eqref{HaCentral} one computes that the component $\cH - \tfrac{1}{4} V \hat{R}$ which mutually commutes with $\hat{R}^*$ satisfies 
\begin{eqnarray} Z\scal{ \cH - \tfrac{1}{4} V \hat{R} } &=& - \frac{3 }{4} e^{-U} Y^2 \N[\Omega]  \; , \CR
 Z_a\scal{ \cH - \tfrac{1}{4} V \hat{R} } &=& \frac{1}{4} e^{-U} \scal{  2 |Y|^2  -   Y^2 }  \Omega_a \; , \end{eqnarray}
which permits to compute that 
\begin{equation} I_4\scal{ \cH - \tfrac{1}{4} V \hat{R} } = 0 \ . \end{equation}
It follows that $ I_4(\cH  - \tfrac{1}{2} V\,\hat R)  $ is linear in $V$, and since the factor $-\frac{1}{2}$ specifically switches the sign of the $\hat{R}$ component one concludes  that 
\begin{equation} e^{-4U} = - I_4(\cH) - M^2 \ . \label{ScallingFactorInvariant} \end{equation}
One obtains in the same way that the moduli can be expressed as 
\begin{equation} t^i = \frac{ - \frac{1}{2} \frac{\partial I_4}{\partial \cH_i}(\cH) + \frac{i}{2} e^{-2U} V \hat{R}^i + ( M + i e^{-2U}) \scal{ \cH^i - \tfrac{i e^{-2U}}{V} \hat{R}^{*i} } }{ - \frac{1}{2} \frac{\partial I_4}{\partial \cH_0}(\cH) + \frac{i}{2} e^{-2U} V \hat{R}^0 + ( M + i e^{-2U}) \scal{ \cH^0 - \tfrac{i e^{-2U}}{V} \hat{R}^{*0} }}  \ , \label{ExplicitModuli}  \end{equation}
after several simplifications.      
The metric is then given by \eqref{metricMltc} with
$\omega$ as in \eqref{omegaM}, whereas the gauge fields are given by
\eqref{gauge-decop} with $\zeta$ as in \eqref{Zetas} and $dw$ given by
\eqref{harmonic-def} above.

\subsection{Physical properties}\label{phys-aspect}

In the preceding subsection, we have treated the constant vector $\hat{R}$ as defining the system, and the constraint \eqref{ZAconstraint} or \eqref{sympl-constr} as a
restriction on the physical charges for a given $\hat{R}$. Physically, it is however more natural to define a solution from the asymptotic moduli $t^i_\infty$, the electromagnetic charge $\Gamma$ and angular momentum $J$. Considering the asymptotic central charge $Z(\Gamma)_\infty$ and its K\"{a}hler derivative $Z_a(\Gamma)_\infty$ (which we will refer to as the `central charges' for simplicity), one can indeed define the asymptotic $\Omega_{a \infty}$ as the unique solution of \eqref{ZAconstraint}  for which $\alpha_\infty$ is determined such that there is no NUT charge, \ie 
\begin{equation} \mbox{Im}\bigl[ \Omega_{a} \bar Z^a(\Gamma) - \N[\Omega] \bar Z(\Gamma) \bigr]_\infty = 0 \ ,\label{NUTzero} \end{equation}
as it is done in \cite{Bossard:2012ge}.  Indeed, it follows from \eqref{ScalLin} that \eqref{NUTzero} defines the NUT charge and that 
\begin{equation} W_\infty = \frac{1}{2} \scal{  \Omega_{a} \bar Z^a(\Gamma) - \N[\Omega] \bar Z(\Gamma) }\big|_\infty = M_{\rm \scriptscriptstyle  ADM} \ ,\label{ADMMass}  \end{equation}
is the non-BPS fake superpotential at spatial infinity, \ie the ADM mass. If one does not fix the gauge for the $U(1)\times K_4$ gauge invariance, the `central charges' do not depend on the flat directions in moduli space, and it follows that $\Omega_{a \infty}$ then does not depend on the flat directions either. However, the constant vectors $\hat{R}$ and $\hat{R}^*$ which are defined from $\Omega_{a \infty}$ upon action of the asymptotic moduli through \eqref{OmegaRhat} and \eqref{Rstar} do.\footnote{Note that $Y_\infty$ is determined by the asymptotic `central charges', because $\alpha_\infty$ is.} Therefore the asymptotic `central charges' altogether with the constant vector $\hat{R}$ contain the information about the flat directions. 

To understand this property, let us discuss the stabilizers of $\hat{R}$ and $\Gamma$ in $G_4$. It is known that the stabilizer of a very small vector as $\hat{R}$ is \cite{Ferrara:2007tu}
\begin{equation} G_5 \ltimes \mathds{R}^{n_v} \subset G_4 \ , \end{equation}
whereas the stabilizer of the electromagnetic charges with a strictly negative quartic invariant $I_4(\Gamma) < 0 $ is $G_5 \subset G_4$.\footnote{This can easily be checked for a D6 very small vector which only non-vanishing charge is $p^0$, because the latter is clearly left invariant by the five-dimensional duality group $G_5$ and the $n_v$ T-dualities. In the same way, a D0-D6 charge with only non-vanishing $q_0$ and $p^0$, is clearly left invariant by $G_5$ only, and admits a negative quartic invariant.} However, as we exhibit in Appendix \ref{app-seed-sol}, only the compact subgroup $K_5 \subset G_5$ of the stabilizer of the charges leaves the very small vector invariant. It follows that the action of the non-compact generators which generate the flat directions 
\begin{equation} G_5 / K_5 \subset G_4 / K_4 \ , \end{equation}
act faithfully on $\hat{R}$. Moreover, one can show that the condition that both $\hat{R}$ and $\hat{R}^*$ are very small, altogether with equation \eqref{sympl-constr}
\begin{equation} \frac12\, I^{'\,M}_{4}(\Gamma,\hat{R}^*) = \Iprod{\hat{R}^*}{\Gamma}\, \Gamma^M
- 2 \frac{\Iprod{\hat{R}^*}{\Gamma}^2 }{\Iprod{\hat{R}^*}{R}} R^M \, , \label{SympCharge}  \end{equation}
entirely determines these small vectors up to an overall rescaling in terms of the electromagnetic charges and $n_v -1$ parameters parametrizing the flat directions. We prove this in Appendix \ref{app-seed-sol} in a specific duality frame. 

Considering a single centre solution carrying charges $\Gamma$ and angular momentum $J$
\begin{equation} \cH = \mathrm{h}  + \frac{\Gamma}{r} \; , \qquad M = \mathrm{m} + J \frac{\cos \theta}{r^2} \; , \end{equation}
the scalar fields on the horizon take the form
\begin{equation}  t^i_* = \frac{ - \frac{1}{2} \frac{\partial I_4}{\partial \Gamma_i}(\Gamma) + 2i S \frac{\Iprod{\hat{R}^*}{\Gamma} \hat{R}^i}{\Iprod{\hat{R}^*}{\hat{R}}}  + \scal{ J \cos \theta + i S} \Scal{ \Gamma^i + i S \frac{\hat{R}^{* i}}{ \Iprod{\hat{R}^*}{\Gamma} } }}{
 - \frac{1}{2} \frac{\partial I_4}{\partial \Gamma_0}(\Gamma) + 2i S \frac{\Iprod{\hat{R}^*}{\Gamma} \hat{R}^0}{\Iprod{\hat{R}^*}{\hat{R}}}   + \scal{ J \cos \theta + i S } \Scal{ \Gamma^0 + i S \frac{\hat{R}^{* 0}}{ \Iprod{\hat{R}^*}{\Gamma} } }}  \ , \label{AttractorModuli}  \end{equation}
 where 
 \begin{equation} S \equiv   \sqrt{ - I_4(\Gamma) - J^2 \cos^2 \theta} \ . \end{equation} 
As is clear from \eqref{AttractorModuli}, the attractor values of the moduli are not entirely determined by the electromagnetic charges $\Gamma$ and the angular momentum $J$, but depend on the asymptotic flat directions through the small vectors $\hat{R}$ and $\hat{R}^*$, in general. This formula \eqref{AttractorModuli} generalises the rotating attractor formula derived in \cite{Astefanesei:2006dd,Bossard:2012ge} in specific duality frames.

Although the scalar fields are not entirely determined by the electromagnetic charges and the angular momentum, it follows from \eqref{ScallingFactorInvariant} that the horizon area only depends on the electromagnetic charges and the angular momentum as expected \cite{Rasheed:1995zv,Larsen:1999pp,Ferrara:2006em,Sen:2005wa}
\begin{equation} A = 4 \pi \sqrt{ - I_4(\Gamma) - J^2 } \ . \label{HA} \end{equation}
The same formula implies that the ADM mass is determined as 
\begin{equation} \label{mass}
M_{\rm \scriptscriptstyle ADM}= - \frac14\,\frac{\partial I_4(\mathrm{h})}{\partial \mathrm{h}^M}\, \Gamma^M
\,,
\end{equation}
where we assumed that asymptotically
\begin{equation}  e^{-4U}|_\infty = -I_4(\mathrm{h}) -\mathrm{m}^2 =1\,. \label{AsympInf} \end{equation}
Although this formula may suggest that the ADM mass is linear in the charges, one must note that its explicit expression in terms the asymptotic `central charges' is generally a non-rational function of the latter \cite{Bossard:2009we}, due to the fact that the integrating constants $\mathrm{h}$ are not entirely parametrized by the asymptotic moduli alone. The situation is similar, although simpler, for BPS black holes, for which the asymptotic moduli are entirely determined in terms of the integrating constants $\mathrm{h}$ of the harmonic functions dual to the electromagnetic vectors, but the reverse is not true, as the constants $\mathrm{h}$ are parametrized by the asymptotic moduli and the phase of the asymptotic central charge. Indeed, the ADM mass of a BPS black hole is not a linear function of the central charge
\begin{equation} M_{\rm \scriptscriptstyle BPS} =  \frac14\,\frac{\partial I_4(\mathrm{h})}{\partial \mathrm{h}^M}\, \Gamma^M = | Z(\Gamma)|_{\infty} \; , \end{equation}
due to the presence of precisely this phase. For the non-BPS solutions, the constants $\rm h$ are also not entirely determined by the asymptotic moduli, since they depend explicitly  on the asymptotic `central charges' through the phase of the central charge as well as the small vectors $\hat{R}$ and $\hat{R}^*$. Moreover, the asymptotic moduli are not entirely determined in terms of the constants $h$ either in this case, as is clear from the asymptotic value of \eqref{ExplicitModuli} 
\begin{equation}  t^i_\infty = \frac{ - \frac{1}{2} \frac{\partial I_4}{\partial {\rm h}_i}({\rm h}) + 2i  \frac{\Iprod{\hat{R}^*}{{\rm h}} \hat{R}^i}{\Iprod{\hat{R}^*}{\hat{R}}}  + \scal{ \mathrm{m} + i } \Scal{ {\rm h}^i + i  \frac{\hat{R}^{* i}}{ \Iprod{\hat{R}^*}{{\rm h}} } }}{
 - \frac{1}{2} \frac{\partial I_4}{\partial {\rm h}_0}({\rm h}) + 2i  \frac{\Iprod{\hat{R}^*}{{\rm h}} \hat{R}^0}{\Iprod{\hat{R}^*}{\hat{R}}}   + \scal{ \mathrm{m} + i  } \Scal{ {\rm h}^0 + i  \frac{\hat{R}^{* 0}}{ \Iprod{\hat{R}^*}{{\rm h}} } }}  \ . \label{AsymModuli}  \end{equation}
This expression includes explicitly the small vectors that depend on the flat directions, and therefore do not affect the mass formula, but includes also the constant $\mathrm{m}$ which parametrizes the phase of $e^{-i\alpha} \N[\Omega] |_{\infty}$, and therefore depends explicitly on the asymptotic `central charges'.

The regularity of the solution requires that $e^{-2U}$ is everywhere strictly positive, and the absence of closed time-like curves outside the horizon moreover requires that $e^{-4U} > \left( \frac{J \sin \theta}{r^2}   \right)^2 $. The latter condition implies the former, and it reads
\begin{equation} - I_4(\cH) > \left(  \mathrm{m} +  \frac{|J|}{r^2} \right)^2  \ . \end{equation}
This condition is clearly satisfied at spatial infinity because of \eqref{AsympInf}, and at the horizon this requires the usual regularity condition
\begin{equation} - I_4(\Gamma) > J^2 \ , \end{equation}
which is necessary for the horizon area \eqref{HA} to be well defined.

\section{Conclusion}
\label{sec:concl}

In this paper we have given a detailed exposition of the first order systems underlying the composite non-BPS system of multi-centre black holes in $\N=2$ supergravity in four dimensions with a symmetric very special K\"{a}hler geometry. Upon imposing a reality constraint on the system of equations, we restricted to the single centre class, which includes all extremal under rotating solutions with one centre and multi-centre generalisations with mutually local charges. Making use of this constraint we were able to explicitly integrate the flow equations for the single centre class for the vector multiplet scalars in a manifestly duality covariant way. Here, we discuss some of the implications of our results.

The solution we obtain for the single centre class, being manifestly duality covariant, allows for general moduli at infinity and arbitrary charge vectors, without the need of dualising a specific seed solution. We stress the presence of an additional (constant) very small vector $\hat R$ and its magnetic dual $\hat{R}^*$ in the solution for the moduli, in addition to the standard vector of harmonic functions describing the charges. This vector arises in the definition of the flow equations for the full composite non-BPS class and therefore plays a central role in our considerations. This is quite different from the squaring of the action in the standard fake superpotential approach for single centre solutions \cite{Ceresole:2007wx, Andrianopoli:2007gt, Perz:2008kh, Ceresole:2009iy, Bossard:2009we, Ceresole:2009vp, Ceresole:2010hq}, which is based on a function of scalars and physical charges only. In view of the fact that our explicit solution \eqref{scal-sol} allows to construct a function driving the flow 
that contains $\hat R$, $\hat{R}^*$ along with the charges and scalars, it remains an interesting open problem to understand the relation between the two formulations.

From a physical point of view both $\hat R$ and $\hat{R}^*$ are integration constants for the scalar equations of motion once the charges are fixed. Indeed, for a single centre solution with given charges it is known that not all scalars take part in the flow from infinity to the near horizon region, but particular combinations are frozen to arbitrary constant values throughout spacetime, the so called flat directions \cite{Tripathy:2005qp, Ferrara:1997uz, Bellucci:2006xz, Ferrara:2007tu, Gimon:2007mh}. We have shown that the ambiguity in defining $\hat R$ from the electromagnetic charges is precisely parametrized by the flat directions in moduli space, as expected in order to describe single centre solutions explicitly. Addition of more centres with charges such that the constraint \eqref{sympl-constr} is satisfied lifts the flat directions, since a unique $\hat R$ is fixed in terms of the charges in the generic case.

The possible microscopic interpretation of the general single centre under-rotating solution remains unclear at the moment. A number of approaches have been proposed for the microscopic construction of extremal non-supersymmetric black holes, see {\it e.g.}~\cite{Emparan:2006it, Dabholkar:2006tb, Gimon:2009gk}. From this point of view, the flat directions appear as geometric moduli or background fluxes that can take arbitrary values, see {\it e.g.}~\cite{Gimon:2007mh, Gimon:2009gk}. 

A natural future direction is the construction of the generic solution in the composite non-BPS class. Since the flow equations are again characterised by the very small vector $\hat R$, it is clear that some of the structures found here will remain relevant in the more general case. The almost BPS class should also admit a similar description.

Finally, the recent results in \cite{Galli:2011fq,Bena:2012wc} indicate that a non-extremal deformation or a lift of our flow equations to five and/or six dimensional supergravity would be very interesting to explore, in connection to the over rotating branch.

\section*{Acknowledgement}
We thank Alessio Marrani for useful discussions. This work was
 supported by the French ANR contract 05-BLAN-NT09-573739, the ERC
Advanced Grant no.  226371, the ITN programme PITN-GA-2009-237920 and
the IFCPAR programme 4104-2.

\begin{appendix}

\section{Under rotating seed solution}\label{app-seed-sol}
In this Appendix we present the known rotating seed solution in a specific
duality
frame \cite{Bena:2009ev}, as a convenient pivot to draw intuition for the
general solution. In this case the electromagnetic vector fields satisfy 
\begin{eqnarray} \star d \tilde{w}_0 &=&  - \frac{1}{\sqrt{2}}  dV\,,  \qquad
\star d w^i = \frac{1}{\sqrt{2}} d L^i\,, \qquad
 d \tilde{w}_i =  d w^0 = 0 \ . 
\end{eqnarray}
The scalar fields take the form
\begin{equation} t^i =\frac{ - M + i e^{-2U}}{\N[L]}\, L^i
\,, \label{seed-scal} \end{equation}
and the metric
\begin{equation} e^{-4U} = V \N[L] - M^2 \ , \qquad \star d \omega  = d M \,.
\label{seed-metric}\end{equation}
In this duality frame the constant small vectors are 
\begin{eqnarray} \hat R=\left( 0,\, 0 \,; \, 2 \sqrt{2} ,\, 0 \right)^T\,, \qquad
    \hat{R}^*=\left(  \sqrt{2} ,\, 0 \,; \, 0,\, 0\right)^T\,,  \label{SpecificRRstar} \end{eqnarray}
which satisfy indeed
\begin{equation}
 Z(\hat{R}^*) =i\,Y^3\,Z(\hat R)\,,
\qquad
 D_iZ(\hat{R}^*) =  -i\, |Y|^2\,Y\,D_iZ(\hat R)\,,
\end{equation}
where we used the definition \eqref{Y-def} and $M$ is the harmonic function in
\eqref{seed-metric}. One can straightforwardly solve \eqref{sympl-constr} in terms of these vectors using that 
\begin{eqnarray} \frac12\, \frac{\partial^2 I_4}{\partial q_0 \partial p^0 } + q_0 p^0 &=&
- q_0 p^0 - q_i p^i  \CR
\frac12\,\frac{\partial^2 I_4}{\partial q_i \partial p^0 } + q_0 p^i &=& -c^{ijk} q_j q_k \CR
\frac12\, \frac{\partial^2 I_4}{\partial p^i \partial p^0 } - q_0 q_i &=& -2\, q_0
q_i   \CR
\frac12\, \frac{\partial^2 I_4}{\partial p^0  \partial p^0 } 
- q_0 q_0 &+& 2 q_0^{\; 2}  = 0 \, .  \label{const-frame} \end{eqnarray}
Because $\cH_0 = - \frac{1}{\sqrt{2}} V$ by definition, the third line implies that $\cH_i = 0 $, and then $\cH^0 = 0$, which was already implied by \eqref{RHzero}. Therefore we find the consistent solution 
\begin{equation} \cH= \frac{1}{\sqrt{2}} \left( 0,\, L^i \,; - V,\, 0\right)^T \ . \end{equation}
It is then straightforward to check that $e^{-4U}$ is indeed equal to \eqref{ScallingFactorInvariant} and that the moduli are equal to \eqref{ExplicitModuli}
\begin{eqnarray} t^i &=& \frac{ ( M + i e^{-2U} ) \frac{1}{\sqrt{2}} L^i }{ - \frac{1}{\sqrt{2}} \N[L] - \frac{i e^{-2U}}{V} ( M + i e^{-2U}) \sqrt{2} } \CR 
&=& - V \frac{ L^i}{M + i e^{-2U} }  =  \frac{ - M + i e^{-2U}}{\N[L]}\, L^i \ . \end{eqnarray}

Let us now consider the general solution of  \eqref{SympCharge} for given charges 
\begin{equation} \Gamma =  \left( 0,\, p^i  \,;  q_0  ,\, 0\right)^T \ , \end{equation}
associated to such a solution. For this purpose, we will parametrize the two small vectors in terms of $n_v+1$ parameters as
\begin{eqnarray} \hat{R}= c \,  \left( - \N[f] ,\, - \tfrac{1}{2} c^{ijk} f_j f_k  \,; \, 1  ,\,- f_i  \right)^T\,, \quad
    \hat{R}^*= c_* \, \left(  1    ,\, e^i   \,; \,  - \N[e] ,\, \tfrac{1}{2} c_{ijk} e^j e^k \right)^T   ,  \label{SmallGenneral} \end{eqnarray}
which can in general be infinite, as long as $\hat{R}$ and $\hat{R}^*$ are themselves finite in the limit. Let us solve  \eqref{SympCharge} as an equation for these two small vectors. The component along $\hat{R}_0$ of this equation implies that 
\begin{equation} \frac{  \scal{ q_0 - \tfrac{1}{2} c_{ijk} p^i e^j e^k }^2}{1 - e^i f_i + \frac{1}{4} c_{ijp} c^{klp} e^i e^j f_k f_l - \N[e]\N[f] } = q_0^{\; 2} \ , \end{equation}
and substituting this into the $\hat{R}_i$ component, one obtains 
\begin{equation} f_i = \frac{1}{q_0} c_{ijk} p^j e^k - \frac{\N[e]}{2 q_0^{\; 2}} c_{ijk} p^j p^k \ . \end{equation}
Substituting these equations and the $\hat{R}^i$ component of \eqref{SympCharge} inside its $\hat{R}^0$ component one finally obtains
\begin{equation} \tfrac{1}{2} c_{ijk} p^i p^j e^k = \frac{1}{q_0} \N[e] \N[p] \ , \end{equation}
which altogether imply that \eqref{SympCharge} is satisfied identically. 

One observes that the small vectors \eqref{SmallGenneral} can then be obtained from the ones in \eqref{SpecificRRstar} by a duality transformation that acts on the scalar fields as
\begin{equation}  t^i(e)  = t^i + e^i + \frac{1}{2 q_0 } c^{ijk} c_{jlp} c_{kqr} e^l p^p t^q t^r - \frac{1}{q_0} t^i t^j c_{jkl} e^k p^l  + \mathcal{O}(e^2) \ , \end{equation}
at first order (if one fixes the irrelevant constants $c=2 \sqrt{2}$ and $c_* = \sqrt{2}$). One straightforwardly computes that such transformations leave the charges $\Gamma$ invariant. In order for the moduli to be well defined, the stabilizer of the charges $p^i$ in $G_5$ must be its maximal compact subgroup $K_5$. Therefore it follows that the $n_v - 1$ $e^i$'s which parametrize the solution of  \eqref{SympCharge}, also parametrize the moduli space of flat directions $G_5 / K_5 \subset G_4 / K_4$, and alternatively, that equation \eqref{sympl-constr} uniquely determines the very small vectors in terms of the electromagnetic charges, up to $n_v-1$ parameters associated to the flat directions.

\section{Duality invariant constraint}\label{app-constraint}
In this appendix we give an outline of the derivation of \eqref{sympl-constr}
from the reality condition (\ref{ZAconstraint}) for a vector $J$ that is
mutually local with the vector $R$, \ie~$\Iprod{J}{R}=0$. To this end, we
compute the vector in \eqref{I4-der}
\begin{equation}  \label{I4-der2}
I^{'\,M}_{4}(J,J,\hat{R}^*)  =
 \frac{1}{2} t^{MNPQ} J_N J_P \hat{R}^*_Q \,,
\end{equation}
where $\hat{R}^*$ the small vector symplectic dual to $R$ defined in \eqref{Rstar}.
Note however that, since the computation is homogeneous with respect to all
vectors, we will rather use
\begin{equation} Z(R^*) = e^{\frac{ 3i \alpha}{2}} \N^\frac{1}{2}[ \bar \Omega] \qquad
Z_a[R^*] = e^{\frac{i\alpha}{2}} \N^\frac{1}{2}[ \bar \Omega]  \Omega_a \,,\end{equation}
here, or in other words we replace $\hat{R^*}$ by its associated vector of mass one.
We now proceed to compute the components of the derivative \eqref{I4-der} in the
complex basis, as in\eqref{I4-cmplx}, starting from the expression
\eqref{I4-def}.
Writing $Z$, $Z_a$ for $Z(J)$, $Z_a(J)$, one obtains that
\begin{multline} \frac{1}{2}   \frac{ \partial I_4}{ \partial \bar Z^a
}(J,J,R^*) = e^{\frac{3i\alpha}{2}} \N^\frac{1}{2}[\bar \Omega] c_{abc} \bar Z^b
\bar Z^c + 2 e^{-\frac{i\alpha}{2}} \N^\frac{1}{2}[\Omega] Z  c_{abc} \bar
\Omega^b \bar Z^c \\  - e^{-\frac{i\alpha}{2}} \N^\frac{1}{2}[\Omega]  c_{abc}
\bar \Omega^b c^{cde} Z_d Z_e - 2 e^{\frac{i\alpha}{2}} \N^\frac{1}{2}[\bar
\Omega]  c_{abc} \bar Z^b c^{cde} \Omega_d Z_e  \\ - e^{\frac{i\alpha}{2}}
\N^\frac{1}{2}[\bar \Omega]  \Omega_a \scal{ Z \bar Z - Z_b \bar Z^b } - 2 Z_a
\mbox{Re} \bigl[ \bar Z e^{\frac{3i\alpha}{2}} \N^\frac{1}{2}[\bar \Omega] -
e^{\frac{i\alpha}{2}} \N^\frac{1}{2}[\bar \Omega]  \Omega_a \bar Z^a \bigr]\,.
\end{multline}
Using the reality constraint  (\ref{ZAconstraint}) one eliminates all the terms
in $ c_{abc} \bar \Omega^b c^{cde} Z_d Z_e $, and using it again on the
resulting expression one eliminates the terms in $c_{abc} \bar Z^b c^{cde}
\Omega_d Z_e$. The final expression is then
\begin{multline} \frac{1}{2}   \frac{ \partial I_4}{ \partial \bar Z^a
}(J,J,R^*) = -2 i Z_a \mbox{Im}  \bigl[ \bar Z e^{\frac{3i\alpha}{2}}
\N^\frac{1}{2}[\bar \Omega] - e^{\frac{i\alpha}{2}} \N^\frac{1}{2}[\bar \Omega]
\Omega_a \bar Z^a \bigr] - \Omega_a \Bigl( e^{\frac{i\alpha}{2}}
\N^\frac{1}{2}[\bar \Omega]  Z \bar Z  \Bigr . \\ \Bigl . + \scal{
e^{\frac{i\alpha}{2}} \N^\frac{1}{2}[\Omega] \bar Z - e^{\frac{i\alpha}{2}}
\N^\frac{1}{2}[\bar \Omega] \Omega_a \bar Z^a + e^{-\frac{i\alpha}{2}}
\N^\frac{1}{2}[\bar \Omega]   Z } \scal{ \N[\bar \Omega] Z - \bar \Omega^a Z_a +
e^{i \alpha} \bar Z + 2 e^{-i \alpha} Z } \Bigr) \,. \label{I4-med}
\end{multline}
In order to simplify the second term we need to use the property that
$\Iprod{J}{R} = 0 $. This permits to solve for
\begin{equation} \label{OmegaZ-inn}
\bar \Omega^a Z_a = \frac{ 3 e^{\frac{-i\alpha}{2}} \N^\frac{1}{2}[\bar
\Omega] Z + e^{\frac{i\alpha}{2}} \N^\frac{1}{2}[\bar \Omega] \scal{ \N[\bar
\Omega] Z + 2 \N[\Omega] \bar Z } }{ e^{\frac{i\alpha}{2}} \N^\frac{1}{2}[\bar
\Omega] + e^{-\frac{i\alpha}{2}} \N^\frac{1}{2}[ \Omega] }\,, \end{equation}
which, when used in \eqref{I4-med} leads to
\begin{multline} \frac{1}{2}   \frac{ \partial I_4}{ \partial \bar Z^a
}(J,J,R^*) = -2 i Z_a \mbox{Im}  \bigl[ \bar Z e^{\frac{3i\alpha}{2}}
\N^\frac{1}{2}[\bar \Omega] - e^{\frac{i\alpha}{2}} \N^\frac{1}{2}[\bar \Omega]
\Omega_a \bar Z^a \bigr] \\ + 4  \Omega_a \mbox{Im}[ e^{-i \alpha } Z ]^2 \frac{
i \mbox{Im}[ e^{-\frac{i\alpha}{2}} \N^\frac{1}{2}[ \Omega] ]}{\mbox{Re}[
e^{\frac{i\alpha}{2}} \N^\frac{1}{2}[\bar \Omega] ]^2}\,. \label{I4A}
\end{multline}
In the same way, one computes the $Z$ component of the derivative
\begin{multline}
 \frac{1}{2}   \frac{ \partial I_4}{ \partial \bar Z }(J,J,R^*) = 2 i Z
\mbox{Im}  \bigl[ \bar Z e^{\frac{3i\alpha}{2}} \N^\frac{1}{2}[\bar \Omega] -
e^{\frac{i\alpha}{2}} \N^\frac{1}{2}[\bar \Omega]  \Omega_a \bar Z^a \bigr] \\
- 4 \N[\Omega]  \mbox{Im}[ e^{-i \alpha } Z ]^2 \frac{ i \mbox{Im}[
e^{-\frac{i\alpha}{2}} \N^\frac{1}{2}[ \Omega] ]}{\mbox{Re}[
e^{\frac{i\alpha}{2}} \N^\frac{1}{2}[\bar \Omega] ]^2}\,.\label{I40}
\end{multline}
In order to interpret these two equations, let us compute that
\begin{eqnarray} \Iprod{R^*}{J} &=& 2 \mbox{Im}\bigl[  e^{-\frac{3i\alpha}{2}}
\N^\frac{1}{2}[ \Omega] Z - e^{-\frac{i\alpha}{2}} \N^\frac{1}{2}[ \Omega] \bar
\Omega^a Z_a \bigr] \CR
&=& - 4 \mbox{Im}[ e^{-i \alpha } Z ] \frac{  \mbox{Im}[ e^{-\frac{i\alpha}{2}}
\N^\frac{1}{2}[ \Omega] ]^2}{\mbox{Re}[ e^{\frac{i\alpha}{2}}
\N^\frac{1}{2}[\bar \Omega] ]^2}\,,
\end{eqnarray}
and
\begin{eqnarray} \Iprod{R^*}{R}  &=& 2 \mbox{Im}\bigl[  e^{-\frac{3i\alpha}{2}}
\N^\frac{1}{2}[ \Omega] \N[\Omega]  - e^{-\frac{i\alpha}{2}} \N^\frac{1}{2}[
\Omega] \bar \Omega^a \Omega_a \bigr] \CR
&=& -8   \mbox{Im}[ e^{-\frac{i\alpha}{2}} \N^\frac{1}{2}[  \Omega] ]^3\,.
\end{eqnarray}
Using these identities we conclude that \eqref{I4A} and \eqref{I40} combine into
\begin{equation}  \frac12\,\frac{ \partial I_4}{ \partial J_M }(J,J,R^*) = \Iprod{R^*}{J} J^M - 2
\frac{\Iprod{R^*}{J}^2 }{\Iprod{R^*}{R}} R^M \,. \end{equation}
Using the homogeneity of this equation in $R$ and $R^*$, one can write it for
the constant vectors $\hat{R}$ and $\hat{R}^*$ such that this equation defines a
quadratic  algebraic equation in $J$, as claimed in section \ref{phys-aspect}.

\end{appendix}

\bibliography{PaperG} \bibliographystyle{JHEP}

\providecommand{\href}[2]{#2}\begingroup\raggedright\begin{thebibliography}{10}

\bibitem{Ferrara:1995ih}
S.~Ferrara, R.~Kallosh, and A.~Strominger, {\it {${\cal N}=2$ extremal black
  holes}},  {\em Phys. Rev.} {\bf D52} (1995) 5412--5416,
  [\href{http://xxx.lanl.gov/abs/hep-th/9508072}{{\tt hep-th/9508072}}].

\bibitem{Strominger:1996kf}
A.~Strominger, {\it {Macroscopic entropy of ${\cal N}=2$ extremal black
  holes}},  {\em Phys. Lett.} {\bf B383} (1996) 39--43,
  [\href{http://xxx.lanl.gov/abs/hep-th/9602111}{{\tt hep-th/9602111}}].

\bibitem{Ferrara:1996dd}
S.~Ferrara and R.~Kallosh, {\it {Supersymmetry and attractors}},  {\em Phys.
  Rev.} {\bf D54} (1996) 1514--1524,
  [\href{http://xxx.lanl.gov/abs/hep-th/9602136}{{\tt hep-th/9602136}}].

\bibitem{Behrndt:1997ny}
K.~Behrndt, D.~{L\"{u}st}, and W.~A. Sabra, {\it {Stationary solutions of
  ${\cal N} = 2$ supergravity}},  {\em Nucl. Phys.} {\bf B510} (1998) 264--288,
  [\href{http://xxx.lanl.gov/abs/hep-th/9705169}{{\tt hep-th/9705169}}].

\bibitem{Denef:2000nb}
F.~Denef, {\it {Supergravity flows and D-brane stability}},  {\em JHEP} {\bf
  08} (2000) 050, [\href{http://xxx.lanl.gov/abs/hep-th/0005049}{{\tt
  hep-th/0005049}}].

\bibitem{LopesCardoso:2000qm}
G.~Lopes~Cardoso, B.~de~Wit, J.~Kappeli, and T.~Mohaupt, {\it {Stationary BPS
  solutions in ${\cal N} = 2$ supergravity with $R^2$ interactions}},  {\em
  JHEP} {\bf 12} (2000) 019,
  [\href{http://xxx.lanl.gov/abs/hep-th/0009234}{{\tt hep-th/0009234}}].

\bibitem{Gauntlett:2002nw}
J.~P. Gauntlett, J.~B. Gutowski, C.~M. Hull, S.~Pakis, and H.~S. Reall, {\it
  {All supersymmetric solutions of minimal supergravity in five dimensions}},
  {\em Class. Quant. Grav.} {\bf 20} (2003) 4587--4634,
  [\href{http://xxx.lanl.gov/abs/hep-th/0209114}{{\tt hep-th/0209114}}].

\bibitem{Gauntlett:2004qy}
J.~P. Gauntlett and J.~B. Gutowski, {\it {General concentric black rings}},
  {\em Phys. Rev.} {\bf D71} (2005) 045002,
  [\href{http://xxx.lanl.gov/abs/hep-th/0408122}{{\tt hep-th/0408122}}].

\bibitem{Castro:2008ne}
A.~Castro, J.~L. Davis, P.~Kraus, and F.~Larsen, {\it {String theory effects on
  five-dimensional black hole Physics}},  {\em Int. J. Mod. Phys.} {\bf A23}
  (2008) 613--691, [\href{http://xxx.lanl.gov/abs/0801.1863}{{\tt
  arXiv:0801.1863}}].

\bibitem{Ceresole:2007wx}
A.~Ceresole and G.~Dall'Agata, {\it {Flow equations for non-BPS extremal black
  holes}},  {\em JHEP} {\bf 03} (2007) 110,
  [\href{http://xxx.lanl.gov/abs/hep-th/0702088}{{\tt hep-th/0702088}}].

\bibitem{LopesCardoso:2007ky}
G.~Lopes~Cardoso, A.~Ceresole, G.~Dall'Agata, J.~M. Oberreuter, and J.~Perz,
  {\it {First-order flow equations for extremal black holes in very special
  geometry}},  {\em JHEP} {\bf 10} (2007) 063,
  [\href{http://xxx.lanl.gov/abs/0706.3373}{{\tt arXiv:0706.3373}}].

\bibitem{Ceresole:2009iy}
A.~Ceresole, G.~Dall'Agata, S.~Ferrara, and A.~Yeranyan, {\it {First order
  flows for ${\cal N}=2$ extremal black holes and duality invariants}},  {\em
  Nucl. Phys.} {\bf B824} (2010) 239--253,
  [\href{http://xxx.lanl.gov/abs/0908.1110}{{\tt arXiv:0908.1110}}].

\bibitem{Gimon:2007mh}
E.~G. Gimon, F.~Larsen, and J.~{Sim\'on}, {\it {Black holes in supergravity:
  the non-BPS branch}},  {\em JHEP} {\bf 01} (2008) 040,
  [\href{http://xxx.lanl.gov/abs/0710.4967}{{\tt arXiv:0710.4967}}].

\bibitem{Gaiotto:2007ag}
D.~Gaiotto, W.~W. Li, and M.~Padi, {\it {Non-supersymmetric attractor flow in
  symmetric spaces}},  {\em JHEP} {\bf 12} (2007) 093,
  [\href{http://xxx.lanl.gov/abs/0710.1638}{{\tt arXiv:0710.1638}}].

\bibitem{Andrianopoli:2007gt}
L.~Andrianopoli, R.~D'Auria, E.~Orazi, and M.~Trigiante, {\it {First order
  description of black holes in moduli space}},  {\em JHEP} {\bf 11} (2007)
  032, [\href{http://xxx.lanl.gov/abs/0706.0712}{{\tt arXiv:0706.0712}}].

\bibitem{Andrianopoli:2009je}
L.~Andrianopoli, R.~D'Auria, E.~Orazi, and M.~Trigiante, {\it {First order
  description of $D=4$ static black holes and the Hamilton--Jacobi equation}},
  {\em Nucl.Phys.} {\bf B833} (2010) 1--16,
  [\href{http://xxx.lanl.gov/abs/0905.3938}{{\tt arXiv:0905.3938}}].

\bibitem{Bossard:2009we}
G.~Bossard, Y.~Michel, and B.~Pioline, {\it {Extremal black holes, nilpotent
  orbits and the true fake superpotential}},  {\em JHEP} {\bf 01} (2010) 038,
  [\href{http://xxx.lanl.gov/abs/0908.1742}{{\tt arXiv:0908.1742}}].

\bibitem{Ceresole:2009vp}
A.~Ceresole, G.~Dall'Agata, S.~Ferrara, and A.~Yeranyan, {\it {Universality of
  the superpotential for $d = 4$ extremal black holes}},  {\em Nucl.Phys.} {\bf
  B832} (2010) 358, [\href{http://xxx.lanl.gov/abs/0910.2697}{{\tt
  arXiv:0910.2697}}].

\bibitem{Perz:2008kh}
J.~Perz, P.~Smyth, T.~Van~Riet, and B.~Vercnocke, {\it {First-order flow
  equations for extremal and non-extremal black holes}},  {\em JHEP} {\bf 03}
  (2009) 150, [\href{http://xxx.lanl.gov/abs/0810.1528}{{\tt
  arXiv:0810.1528}}].

\bibitem{Kim:2010bf}
S.-S. Kim, J.~Lindman~{H\"{o}rnlund}, J.~Palmkvist, and A.~Virmani, {\it
  {Extremal solutions of the $S^3$ model and nilpotent orbits of $G_{2(2)}$}},
  {\em JHEP} {\bf 1008} (2010) 072,
  [\href{http://xxx.lanl.gov/abs/1004.5242}{{\tt arXiv:1004.5242}}].

\bibitem{Galli:2010mg}
P.~Galli, K.~Goldstein, S.~Katmadas, and J.~Perz, {\it {First-order flows and
  stabilisation equations for non-BPS extremal black holes}},  {\em JHEP} {\bf
  1106} (2011) 070, [\href{http://xxx.lanl.gov/abs/1012.4020}{{\tt
  arXiv:1012.4020}}].

\bibitem{Rasheed:1995zv}
D.~Rasheed, {\it {The rotating dyonic black holes of Kaluza--Klein theory}},
  {\em Nucl. Phys.} {\bf B454} (1995) 379--401,
  [\href{http://xxx.lanl.gov/abs/hep-th/9505038}{{\tt hep-th/9505038}}].

\bibitem{Matos:1996km}
T.~Matos and C.~Mora, {\it {Stationary dilatons with arbitrary electromagnetic
  field}},  {\em Class. Quant. Grav.} {\bf 14} (1997) 2331--2340,
  [\href{http://xxx.lanl.gov/abs/hep-th/9610013}{{\tt hep-th/9610013}}].

\bibitem{Larsen:1999pp}
F.~Larsen, {\it {Rotating Kaluza--Klein black holes}},  {\em Nucl. Phys.} {\bf
  B575} (2000) 211--230, [\href{http://xxx.lanl.gov/abs/hep-th/9909102}{{\tt
  hep-th/9909102}}].

\bibitem{Ortin:1996bz}
T.~{Ort\'{\i}n}, {\it {Extremality versus supersymmetry in stringy black
  holes}},  {\em Phys. Lett.} {\bf B422} (1998) 93--100,
  [\href{http://xxx.lanl.gov/abs/hep-th/9612142}{{\tt hep-th/9612142}}].

\bibitem{Khuri:1995xq}
R.~R. Khuri and T.~{Ort\'{\i}n}, {\it {A Nonsupersymmetric dyonic extreme
  Reissner-Nordstrom black hole}},  {\em Phys.Lett.} {\bf B373} (1996) 56--60,
  [\href{http://xxx.lanl.gov/abs/hep-th/9512178}{{\tt hep-th/9512178}}].

\bibitem{Tripathy:2005qp}
P.~K. Tripathy and S.~P. Trivedi, {\it {Non-supersymmetric attractors in string
  theory}},  {\em JHEP} {\bf 03} (2006) 022,
  [\href{http://xxx.lanl.gov/abs/hep-th/0511117}{{\tt hep-th/0511117}}].

\bibitem{Astefanesei:2006dd}
D.~Astefanesei, K.~Goldstein, R.~P. Jena, A.~Sen, and S.~P. Trivedi, {\it
  {Rotating attractors}},  {\em JHEP} {\bf 10} (2006) 058,
  [\href{http://xxx.lanl.gov/abs/hep-th/0606244}{{\tt hep-th/0606244}}].

\bibitem{Bena:2009ev}
I.~Bena, G.~Dall'Agata, S.~Giusto, C.~Ruef, and N.~P. Warner, {\it {Non-BPS
  black rings and black holes in Taub-NUT}},  {\em JHEP} {\bf 06} (2009) 015,
  [\href{http://xxx.lanl.gov/abs/0902.4526}{{\tt arXiv:0902.4526}}].

\bibitem{Dall'Agata:2010dy}
G.~Dall'Agata, S.~Giusto, and C.~Ruef, {\it {U-duality and non-BPS solutions}},
   {\em JHEP} {\bf 1102} (2011) 074,
  [\href{http://xxx.lanl.gov/abs/1012.4803}{{\tt arXiv:1012.4803}}].

\bibitem{Goldstein:2008fq}
K.~Goldstein and S.~Katmadas, {\it {Almost BPS black holes}},  {\em JHEP} {\bf
  05} (2009) 058, [\href{http://xxx.lanl.gov/abs/0812.4183}{{\tt
  arXiv:0812.4183}}].

\bibitem{Bossard:2011kz}
G.~Bossard and C.~Ruef, {\it {Interacting non-BPS black holes}},  {\em
  Gen.Rel.Grav.} {\bf 44} (2012) 21--66,
  [\href{http://xxx.lanl.gov/abs/1106.5806}{{\tt arXiv:1106.5806}}].

\bibitem{Bena:2009en}
I.~Bena, S.~Giusto, C.~Ruef, and N.~P. Warner, {\it {Multi-center non-BPS black
  holes - the solution}},  {\em JHEP} {\bf 11} (2009) 032,
  [\href{http://xxx.lanl.gov/abs/0908.2121}{{\tt arXiv:0908.2121}}].

\bibitem{Bena:2009qv}
I.~Bena, S.~Giusto, C.~Ruef, and N.~P. Warner, {\it {A (running) bolt for new
  reasons}},  {\em JHEP} {\bf 11} (2009) 089,
  [\href{http://xxx.lanl.gov/abs/0909.2559}{{\tt arXiv:0909.2559}}].

\bibitem{Bena:2009fi}
I.~Bena, S.~Giusto, C.~Ruef, and N.~P. Warner, {\it {Supergravity solutions
  from floating branes}},  {\em JHEP} {\bf 03} (2010) 047,
  [\href{http://xxx.lanl.gov/abs/0910.1860}{{\tt arXiv:0910.1860}}].

\bibitem{Bobev:2009kn}
N.~Bobev and C.~Ruef, {\it {The Nuts and Bolts of Einstein--Maxwell
  Solutions}},  {\em JHEP} {\bf 1001} (2010) 124,
  [\href{http://xxx.lanl.gov/abs/0912.0010}{{\tt arXiv:0912.0010}}].

\bibitem{Breitenlohner:1987dg}
P.~Breitenlohner, D.~Maison, and G.~W. Gibbons, {\it {Four-dimensional black
  holes from Kaluza--Klein theories}},  {\em Commun. Math. Phys.} {\bf 120}
  (1988) 295.

\bibitem{Gunaydin:2005mx}
M.~{G\"{u}naydin}, A.~Neitzke, B.~Pioline, and A.~Waldron, {\it {BPS black
  holes, quantum attractor flows and automorphic forms}},  {\em Phys.Rev.} {\bf
  D73} (2006) 084019, [\href{http://xxx.lanl.gov/abs/hep-th/0512296}{{\tt
  hep-th/0512296}}].

\bibitem{Bergshoeff:2008be}
E.~Bergshoeff, W.~Chemissany, A.~Ploegh, M.~Trigiante, and T.~Van~Riet, {\it
  {Generating geodesic flows and supergravity solutions}},  {\em Nucl.Phys.}
  {\bf B812} (2009) 343--401, [\href{http://xxx.lanl.gov/abs/0806.2310}{{\tt
  arXiv:0806.2310}}].

\bibitem{Bossard:2009at}
G.~Bossard, H.~Nicolai, and K.~S. Stelle, {\it {Universal BPS structure of
  stationary supergravity solutions}},  {\em JHEP} {\bf 07} (2009) 003,
  [\href{http://xxx.lanl.gov/abs/0902.4438}{{\tt arXiv:0902.4438}}].

\bibitem{Bossard:2009my}
G.~Bossard and H.~Nicolai, {\it {Multi-black holes from nilpotent Lie algebra
  orbits}},  {\em Gen. Rel. Grav.} {\bf 42} (2010) 509--537,
  [\href{http://xxx.lanl.gov/abs/0906.1987}{{\tt arXiv:0906.1987}}].

\bibitem{Fre:2011uy}
P.~{Fr\'e}, A.~S. Sorin, and M.~Trigiante, {\it {Integrability of supergravity
  black holes and new tensor classifiers of regular and nilpotent orbits}},
  {\em JHEP} {\bf 1204} (2012) 015,
  [\href{http://xxx.lanl.gov/abs/1103.0848}{{\tt arXiv:1103.0848}}].

\bibitem{Chemissany:2012nb}
W.~Chemissany, P.~Giaccone, D.~Ruggeri, and M.~Trigiante, {\it {Black hole
  solutions to the $F_4$-model and their orbits (I)}},
  \href{http://xxx.lanl.gov/abs/1203.6338}{{\tt arXiv:1203.6338}}.

\bibitem{Bossard:2012ge}
G.~Bossard, {\it {Octonionic black holes}},
  \href{http://xxx.lanl.gov/abs/1203.0530}{{\tt arXiv:1203.0530}}.

\bibitem{Bossard:2010mv}
G.~Bossard, {\it {1/8 BPS black hole composites}},
  \href{http://xxx.lanl.gov/abs/1001.3157}{{\tt arXiv:1001.3157}}.

\bibitem{Kallosh:2006ib}
R.~Kallosh, N.~Sivanandam, and M.~Soroush, {\it {Exact attractive non-BPS STU
  black holes}},  {\em Phys. Rev.} {\bf D74} (2006) 065008,
  [\href{http://xxx.lanl.gov/abs/hep-th/0606263}{{\tt hep-th/0606263}}].

\bibitem{deWit:1984pk}
B.~de~Wit and A.~Van~Proeyen, {\it {Potentials and symmetries of general gauged
  ${\cal N}=2$ supergravity: Yang--Mills models}},  {\em Nucl. Phys.} {\bf
  B245} (1984) 89.

\bibitem{deWit:1984px}
B.~de~Wit, P.~G. Lauwers, and A.~Van~Proeyen, {\it {Lagrangians of ${\cal N}=2$
  supergravity - matter systems}},  {\em Nucl. Phys.} {\bf B255} (1985) 569.

\bibitem{Ceresole:1995ca}
A.~Ceresole, R.~D'Auria, and S.~Ferrara, {\it {The symplectic structure of
  ${\cal N}=2$ supergravity and its central extension}},  {\em Nucl. Phys.
  Proc. Suppl.} {\bf 46} (1996) 67--74,
  [\href{http://xxx.lanl.gov/abs/hep-th/9509160}{{\tt hep-th/9509160}}].

\bibitem{Gunaydin:1983bi}
M.~{G\"{u}naydin}, G.~Sierra, and P.~Townsend, {\it {The geometry of ${\cal
  N}=2$ Maxwell--Einstein supergravity and Jordan algebras}},  {\em Nucl.Phys.}
  {\bf B242} (1984) 244.

\bibitem{Ferrara:1989ik}
S.~Ferrara and S.~Sabharwal, {\it {Quaternionic manifolds for type II
  superstring vacua of Calabi--Yau spaces}},  {\em Nucl.Phys.} {\bf B332}
  (1990) 317.

\bibitem{Ferrara:2006yb}
S.~Ferrara, E.~G. Gimon, and R.~Kallosh, {\it {Magic supergravities, ${\cal N}=
  8$ and black hole composites}},  {\em Phys.Rev.} {\bf D74} (2006) 125018,
  [\href{http://xxx.lanl.gov/abs/hep-th/0606211}{{\tt hep-th/0606211}}].

\bibitem{Bates:2003vx}
B.~Bates and F.~Denef, {\it {Exact solutions for supersymmetric stationary
  black hole composites}},  {\em JHEP} {\bf 1111} (2011) 127,
  [\href{http://xxx.lanl.gov/abs/hep-th/0304094}{{\tt hep-th/0304094}}].

\bibitem{Ferrara:2010ug}
S.~Ferrara, A.~Marrani, E.~Orazi, R.~Stora, and A.~Yeranyan, {\it {Two-center
  black holes duality-invariants for STU model and its lower-rank
  descendants}},  \href{http://xxx.lanl.gov/abs/1011.5864}{{\tt
  arXiv:1011.5864}}.

\bibitem{Ferrara:2007tu}
S.~Ferrara and A.~Marrani, {\it {On the moduli space of non-BPS attractors for
  ${\cal N}=2$ symmetric manifolds}},  {\em Phys. Lett.} {\bf B652} (2007)
  111--117, [\href{http://xxx.lanl.gov/abs/0706.1667}{{\tt arXiv:0706.1667}}].

\bibitem{Ferrara:2006em}
S.~Ferrara and R.~Kallosh, {\it {On ${\cal N}=8$ attractors}},  {\em Phys.Rev.}
  {\bf D73} (2006) 125005, [\href{http://xxx.lanl.gov/abs/hep-th/0603247}{{\tt
  hep-th/0603247}}].

\bibitem{Sen:2005wa}
A.~Sen, {\it {Black hole entropy function and the attractor mechanism in higher
  derivative gravity}},  {\em JHEP} {\bf 09} (2005) 038,
  [\href{http://xxx.lanl.gov/abs/hep-th/0506177}{{\tt hep-th/0506177}}].

\bibitem{Ceresole:2010hq}
A.~Ceresole and S.~Ferrara, {\it {Black holes and attractors in supergravity}},
   \href{http://xxx.lanl.gov/abs/1009.4175}{{\tt arXiv:1009.4175}}.

\bibitem{Ferrara:1997uz}
S.~Ferrara and M.~{G\"{u}naydin}, {\it {Orbits of exceptional groups, duality
  and BPS states in string theory}},  {\em Int. J. Mod. Phys.} {\bf A13} (1998)
  2075--2088, [\href{http://xxx.lanl.gov/abs/hep-th/9708025}{{\tt
  hep-th/9708025}}].

\bibitem{Bellucci:2006xz}
S.~Bellucci, S.~Ferrara, M.~{G\"{u}naydin}, and A.~Marrani, {\it {Charge orbits
  of symmetric special geometries and attractors}},  {\em Int. J. Mod. Phys.}
  {\bf A21} (2006) 5043--5098,
  [\href{http://xxx.lanl.gov/abs/hep-th/0606209}{{\tt hep-th/0606209}}].

\bibitem{Emparan:2006it}
R.~Emparan and G.~T. Horowitz, {\it {Microstates of a neutral black hole in M
  theory}},  {\em Phys.Rev.Lett.} {\bf 97} (2006) 141601,
  [\href{http://xxx.lanl.gov/abs/hep-th/0607023}{{\tt hep-th/0607023}}].

\bibitem{Dabholkar:2006tb}
A.~Dabholkar, A.~Sen, and S.~P. Trivedi, {\it {Black hole microstates and
  attractor without supersymmetry}},  {\em JHEP} {\bf 0701} (2007) 096,
  [\href{http://xxx.lanl.gov/abs/hep-th/0611143}{{\tt hep-th/0611143}}].

\bibitem{Gimon:2009gk}
E.~G. Gimon, F.~Larsen, and J.~{Sim\'on}, {\it {Constituent model of extremal
  non-BPS black holes}},  {\em JHEP} {\bf 07} (2009) 052,
  [\href{http://xxx.lanl.gov/abs/0903.0719}{{\tt arXiv:0903.0719}}].

\bibitem{Galli:2011fq}
P.~Galli, T.~{Ort\'{\i}n}, J.~Perz, and C.~S. Shahbazi, {\it {Non-extremal
  black holes of ${\cal N}=2$, $d=4$ supergravity}},  {\em JHEP} {\bf 1107}
  (2011) 041, [\href{http://xxx.lanl.gov/abs/1105.3311}{{\tt
  arXiv:1105.3311}}].

\bibitem{Bena:2012wc}
I.~Bena, M.~Guica, and W.~Song, {\it {Un-twisting the NHEK with spectral
  flows}},  \href{http://xxx.lanl.gov/abs/1203.4227}{{\tt arXiv:1203.4227}}.

\end{thebibliography}\endgroup
\end{document}